\documentclass{aastex}
\usepackage{amsmath}
\usepackage{graphicx}
\usepackage{natbib}
\usepackage{multirow}

\shorttitle{Fallback Mechanisms in Core-Collapse Supernovae}
\shortauthors{T.-W. Wong et al.}

\begin{document}

\title{The Fallback Mechanisms in Core-Collapse Supernovae}
\author{
Tsing-Wai Wong$^{1,2}$,
Christopher L. Fryer$^3$,
Carola I. Ellinger$^4$,
Gabriel Rockefeller$^3$,
Vassiliki Kalogera$^1$
}

\affil{
$^1$Center for Interdisciplinary Exploration and Research in Astrophysics (CIERA) \& Department of Physics and Astronomy, Northwestern University,
\\2145 Sheridan Road, Evanston, IL 60208, USA; tsingwong2012@u.northwestern.edu
\\$^2$Harvard-Smithsonian Center for Astrophysics, 60 Garden St., Cambridge, MA 02138, USA
\\$^3$CCS-2, MS D409, Los Alamos National Laboratory, Los Alamos, NM, USA
\\$^4$Department of Physics, University of Texas at Arlington, 502 Yates Street, Box 19059, Arlington, TX 76019, USA
}

\begin{abstract}
Although the details of the core-collapse supernova mechanism are not fully understood, it is generally accepted that the energy released in the collapse
produces a shock that disrupts the star and produces the explosion.
Some of the stellar material does not receive enough energy to escape the potential well of the newly formed neutron star and it falls back on to the core.
This fallback plays an important role in setting compact remnant masses and has been invoked to explain a number of observed phenomena:
late-time neutrino emission, r-process element production, peculiar supernovae, and long-duration gamma-ray bursts.
To examine the link between fallback and these observed phenomena, we must understand the nature of fallback and currently, multiple mechanisms have
been proposed that lead to different characteristics of the fallback.
In this paper, we model a series of three-dimensional explosions by following the fallback for two different massive progenitors to study both the nature of
the fallback and the role of asymmetric explosions on this fallback.
One of the most-cited mechanisms behind fallback argues that it is caused by the reverse shock produced when the outward moving supernova shock
decelerates in the relatively flat density profile of the stellar envelope.
We show that this mechanism plays a minor, if any, role in driving fallback and most supernova fallback is produced by prompt mechanisms.
\end{abstract}

\section{Introduction}
Fallback in supernovae was first proposed to prevent the ejection of neutron rich material that was not seen in observations of nucleosynthetic
yields \citep{colgate71}.
Since that time, most (if not all) supernova explosions have exhibited some amount of fallback, while the most solid evidence for fallback has waited
for accurate mass estimates of compact remnants.

Over the past few decades, many compact remnants were found in our Galaxy and other galaxies in the local universe.
The masses of compact remnants are well constrained if the remnant is in a binary system.
Hence, these objects are the focus of mass distribution studies of compact remnants
\citep[e.g.][]{finn94, bailyn...98, thorsett=chakrabarty99, kaper...06, nice...08, ozel...10, ozel...12, schwab...10, farr...11, kreidberg...12, kiziltan...13}.
Here, we briefly summarize the results from the most recent studies.
\cite{ozel...12} considered neutron stars observed in all different kinds of binaries and argued that neutron stars have bimodal mass distribution:
$1.28 \pm 0.24$\,$M_\odot$ for non-recycled pulsars and  $1.48 \pm 0.2$\,$M_\odot$ for recycled pulsars.
Using neutron stars with neutron star or white dwarf companions, \cite{kiziltan...13} also found a bimodal mass distribution,
with respective peaks at 1.33 and 1.55\,$M_\odot$.
This bimodal mass distribution can be explained by mass accretion onto the neutron star due to binary interactions \citep[e.g.][]{ozel...12, kiziltan...13}.
At the present time, the most massive neutron star is observed to be $\sim$2\,$M_\odot$ \citep{freire...08, demorest...10}.
For (stellar) black holes, \cite{ozel...10} considered the masses of the black holes in low-mass X-ray binaries and derived a mass distribution
of $7.8 \pm 1.2$\,$M_\odot$.
\cite{farr...11} also studied the masses of black holes in low-mass X-ray binaries and argued that the minimum black hole mass falls in the range of
4.3--4.5\,$M_\odot$ at 90\% confidence level.
These results imply that there is a significant gap between the masses of the observed black holes and neutron stars.
\cite{kreidberg...12} suggested that this mass gap could be the result of systematic errors in the orbital inclination angles of the observed black hole
X-ray binaries, which were used in determining the mass functions of those black holes.
Nevertheless, even with this bias, a paucity of low-mass black holes remains.
Such a mass gap or paucity can provide constraints on the physics of core-collapse supernova \citep[see e.g.][]{fryer99, fryer12, belczynski12, kochanek13}.

Previous studies of core-collapse supernovae have demonstrated the importance of fallback in determining the mass distribution of compact remnants,
and have showed that the mass of a compact remnant depends on both the stellar progenitor and the supernova explosion energy produced in the
collapse of the stellar core
\citep{fryer99, fryerkalogera01, macfadyen01, fryeretal06, young07, zhang08, fryer09, fragos...09, oconner11, fryer12, ugliano12, dexter13}.
Early estimates of the mass of the compact remnant in stellar modeling neglected this fallback and used the structure of the stellar core
at collapse to define the remnant mass.
For example, \cite{timmes96} set the remnant mass equal to the mass of the iron core at collapse,
producing a double, nearly delta function peak in the distribution of the remnant masses.
More sophisticated estimates of the remnant mass coupled the progenitor structure with the explosive mechanism,
broadening the range of remnant masses to some extent \citep{fryerkalogera01,oconner11,fryer12}.
However, fallback is required to produce the observed broad range of compact remnant masses.

At the present time, there are three broad mechanisms invoked to explain fallback:
rarefaction wave deceleration, energy and momentum loss of the ejecta, and reverse shock deceleration.
In this paper, ``ejecta'' not only denotes the supernova remnant, but also the material behind the forward shock before it breaks through the stellar surface.
For the reason explained in what follows, we group the first two fallback mechanisms as ``prompt fallback mechanism''.
After the launch of the forward shock, the proto-neutron star cools and contracts, sending a rarefaction wave toward the ejecta.
This rarefaction wave decelerates part of the ejecta sufficiently that it no longer has enough energy to escape from the gravity of the proto-neutron star,
and eventually falls back \citep{colgate71}.
Associated with this fallback is the fact that the ejecta loses energy (PdV work) and momentum as it pushes outward, exploding the star.
Due to this energy and momentum loss, part of the ejecta is decelerated below the local escape velocity, causing it to fall back onto the
proto-neutron star \citep{fryer99}.
These two effects combine to produce fallback at early times and we will refer to this scenario as the ``prompt fallback mechanism''.

Alternatively, by focusing on the velocity of the forward shock itself (not the ejecta in total), another mechanism arises.
The \cite{sedov59} solution for a blast wave argues that the velocity of the forward shock ($v_{\rm shock}$) is proportional to
$t^{(\omega-3)/(5-\omega)}$, where $t$ is the time and $\omega$ is defined by the the density profile $\rho(r) \propto r^{-\omega}$.
When the forward shock hits the hydrogen layer of the star, the density profile flattens out and the forward shock decelerates rapidly,
driving a reverse shock through the ejecta.
This reverse shock decelerates the ejecta, leading to fallback \citep{chevalier89, woosley89, zhang08}.
The primary difference between the prompt fallback mechanism and the reverse shock mechanism is the timeframe for fallback.
Typically, the prompt fallback mechanism drives fallback quickly \citep[first $\sim 15$\,s,][]{fryer09},
whereas the reverse shock mechanism must wait for the forward shock to reach the hydrogen layer (typically a few hundred seconds)
before the reverse shock is produced, leading to fallback occurring a few hundred to a thousand seconds after the launch of the explosion.

Simulations of fallback can be divided into two broad classes.
One class of simulations varies both the progenitor mass and explosion energy to study the nature of fallback
\citep{macfadyen01, young07, zhang08, ugliano12, dexter13},
The other class uses an understanding, albeit imperfect, of the explosion mechanism to estimate the energy, and hence fallback, for a particular progenitor:
e.g. in the case of the convective engine, the energy stored in the turbulent region is roughly the energy available for the supernova
explosion \citep{fryer99, fryerkalogera01, fryer12}.
The latter class are useful in studying remnant mass populations \citep[e.g.][]{fryerkalogera01, fryer12, belczynski12},
while the former class are ideally suited to studying the nature of fallback:
e.g. which mechanism (prompt fallback or reverse shock) drives fallback and fallback decay rates and timescales.
In this study, we will focus on these aspects in order to study the nature of fallback, hence we will adopt the approach taken by the former class of fallback
simulations.

Calculations of fallback face a series of challenges.
First of all, there are uncertainties in the stellar models of supernova immediate progenitors \citep[see e.g.][]{young=arnett05, langer12}.
Their structure depend on the strength of mixing due to rotation \citep{maeder=meynet00, maeder09book},
convective overshooting \citep{stothers=chin85, langer91, herwig...97} and semi-convection \citep{chiosi=summa70, langer91, yoon...06}.
It is well known that the evolution of massive stars are sensitive to mass loss \citep[e.g.][]{chiosi...78, chiosi=maeder86}.
However, the mass-loss rates of massive stars are still not well constrained, especially for the evolution after the main-sequence phase
\citep[e.g.][]{mokiem...07, moffat08, debeck...10, grafener...11, mauron=josselin11, vink...11}.

Beyond uncertainties in the supernova progenitors, different explosion-driven engines and numerical uncertainties,
particularly in the treatment of the inner boundary, can alter the amount of fallback.
Simulations that use piston models to drive an explosion \citep{woosley89, woosley=weaver95, macfadyen01, zhang08, dexter13} are dominated by reverse shock accretion,
because the piston itself prevents early-time fallback.
The fallback from piston engines is typically much lower than those from the more physically justified energy- or entropy-driven engines \citep{young07},
which deposit energy or entropy above the proto-neutron star surface until an explosion is launched.
The piston engine matches only those engines that continue to drive after the launch of the explosion.
The fallback from energy-driven engines is typically dominated by prompt fallback \citep{macfadyen01, young07, fryer09}.
Careful energy deposition and inner boundary treatment (e.g. position of boundary, hard or absorptive boundary, and neutrino cooling at the boundary surface)
are crucial to obtain accurate fallback estimates \citep{young07}.
In addition, aside from a few cases \citep[e.g.][]{fryer=young07, fragos...09, bruenn...13, ellinger...13, hanke...13}, most of these fallback calculations are done
in one dimension (1D).
Multi-dimensional effects must be included to model fallback accurately.

In this paper, we use three-dimensional (3D) hydrodynamic simulations to study two mechanisms leading to fallback: prompt fallback 
and reverse shock deceleration.
We study both symmetric and asymmetric explosions, and we find that prompt fallback is the dominant mechanism across a wide range of explosion
energies and geometries.
The methodology involved in our simulations are discussed in \S\,\ref{sec:methodology}, while the results are shown in \S\,\ref{sec:result}.
We expand on our conclusions and discussion in the final section.

\section{Explosion Simulations}
\label{sec:methodology}

\subsection{Initial Progenitors and Collapse Calculation}
\label{sec:1Dmodel}

\begin{figure}
\begin{center}
\plottwo{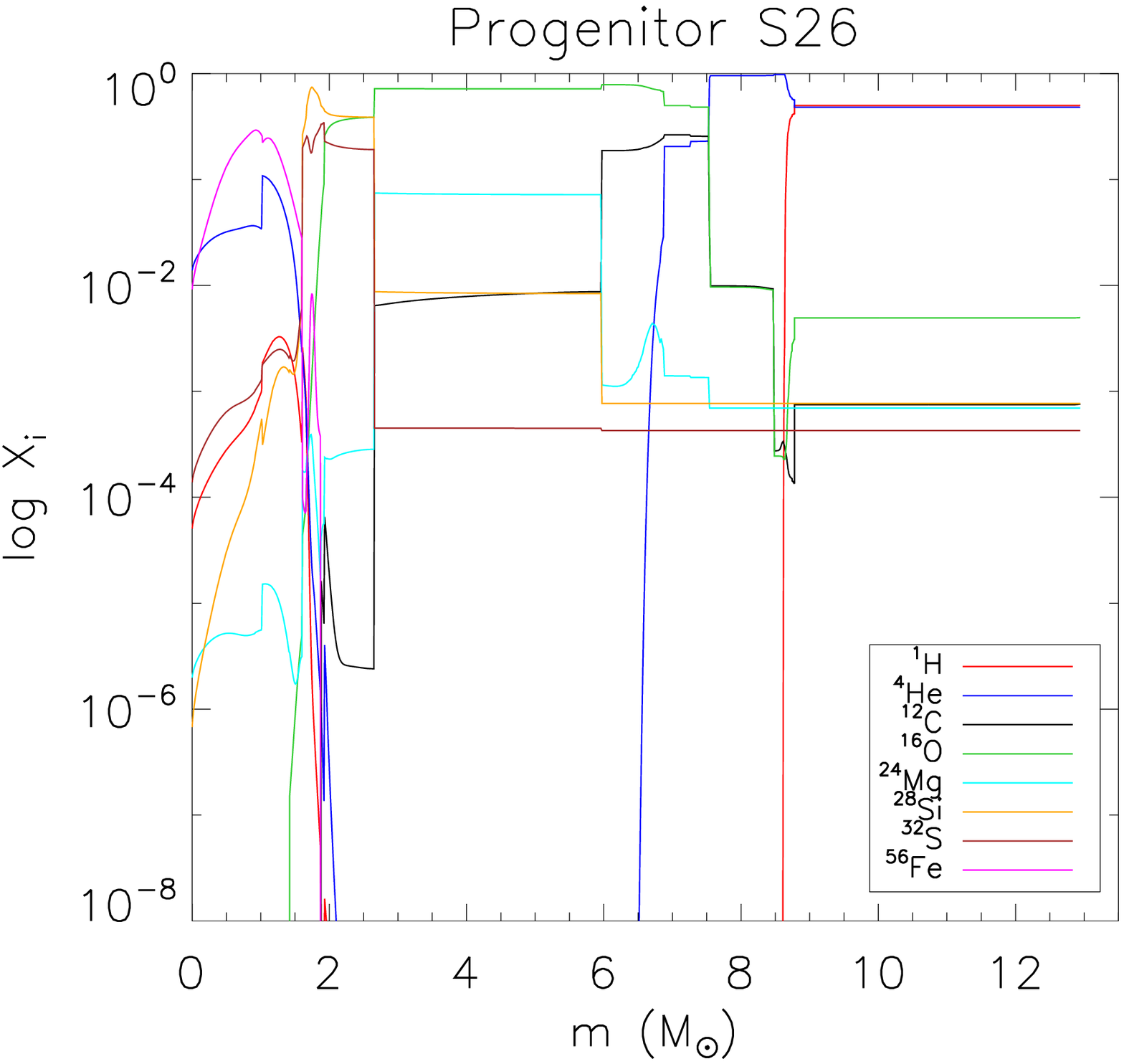}{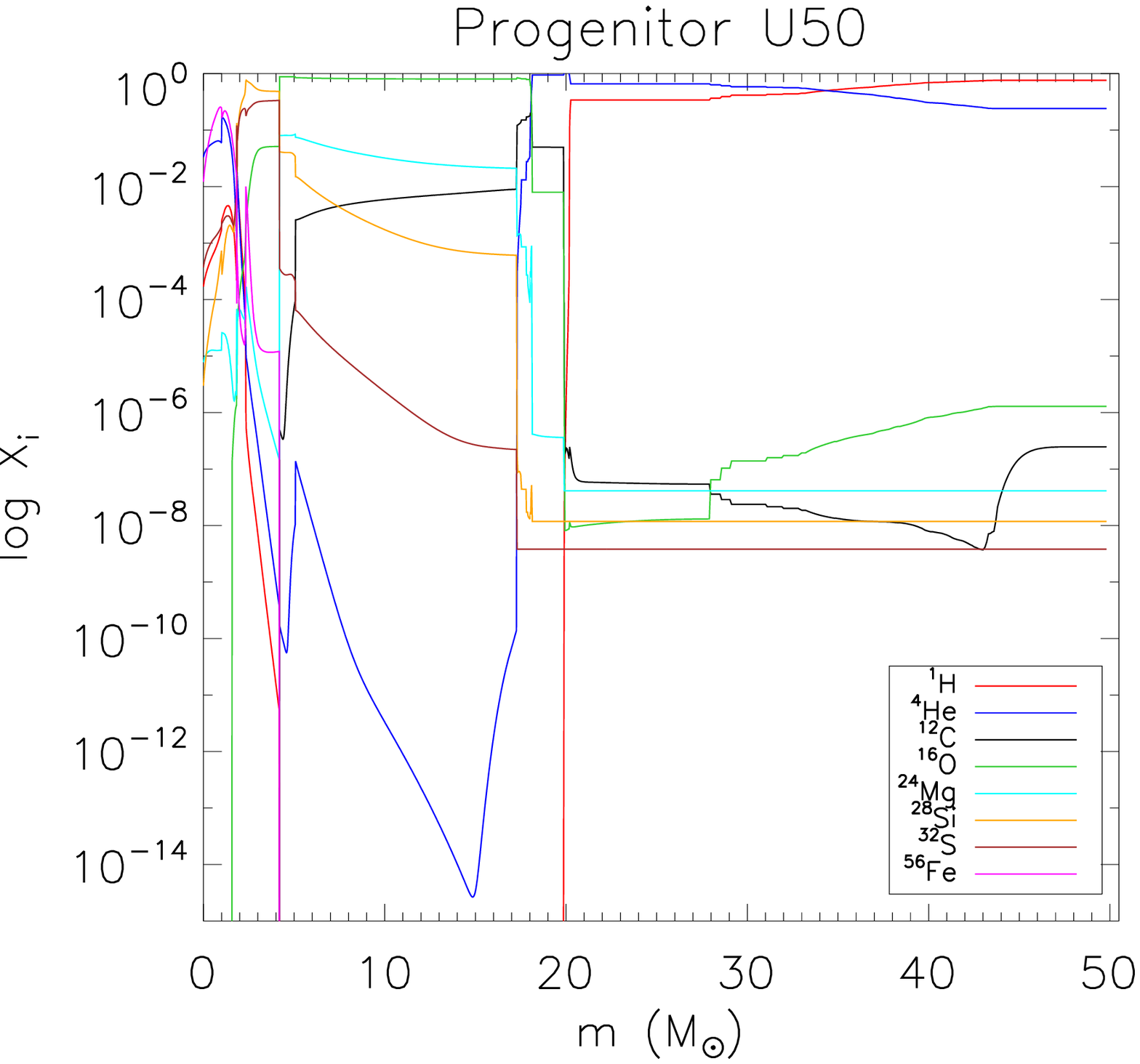}
\caption{The abundance profiles of the progenitor S26 and U50 at the beginning of core collapse.
$X_i$ is the mass fraction of the isotope $i$ given in the legend.}
\label{fig:abund}
\end{center}
\end{figure}

We consider two progenitor star models of \cite{woosley...02}: a 26~$M_\odot$ star of solar metallicity and a 50~$M_\odot$ star of zero metallicity.
Hereafter, we refer to the former and latter progenitor model as S26 and U50, respectively.
Our progenitors are non-rotating stars evolved with the stellar evolutionary code KEPLER \citep{weaver...78, heger...00, woosley...02}.
Their evolution accounts for mass loss via stellar wind and mixing of elements due to convection and overshooting.
At core collapse, progenitor S26 evolves to a 13\,$M_\odot$ supergiant with a radius of 1500\,$R_\odot$,
while progenitor U50 has only lost a negligible amount of mass.
The abundance profiles of both progenitors at core collapse are shown in Figure~\ref{fig:abund}.

\begin{figure}
\begin{center}
\includegraphics[width=5.4cm]{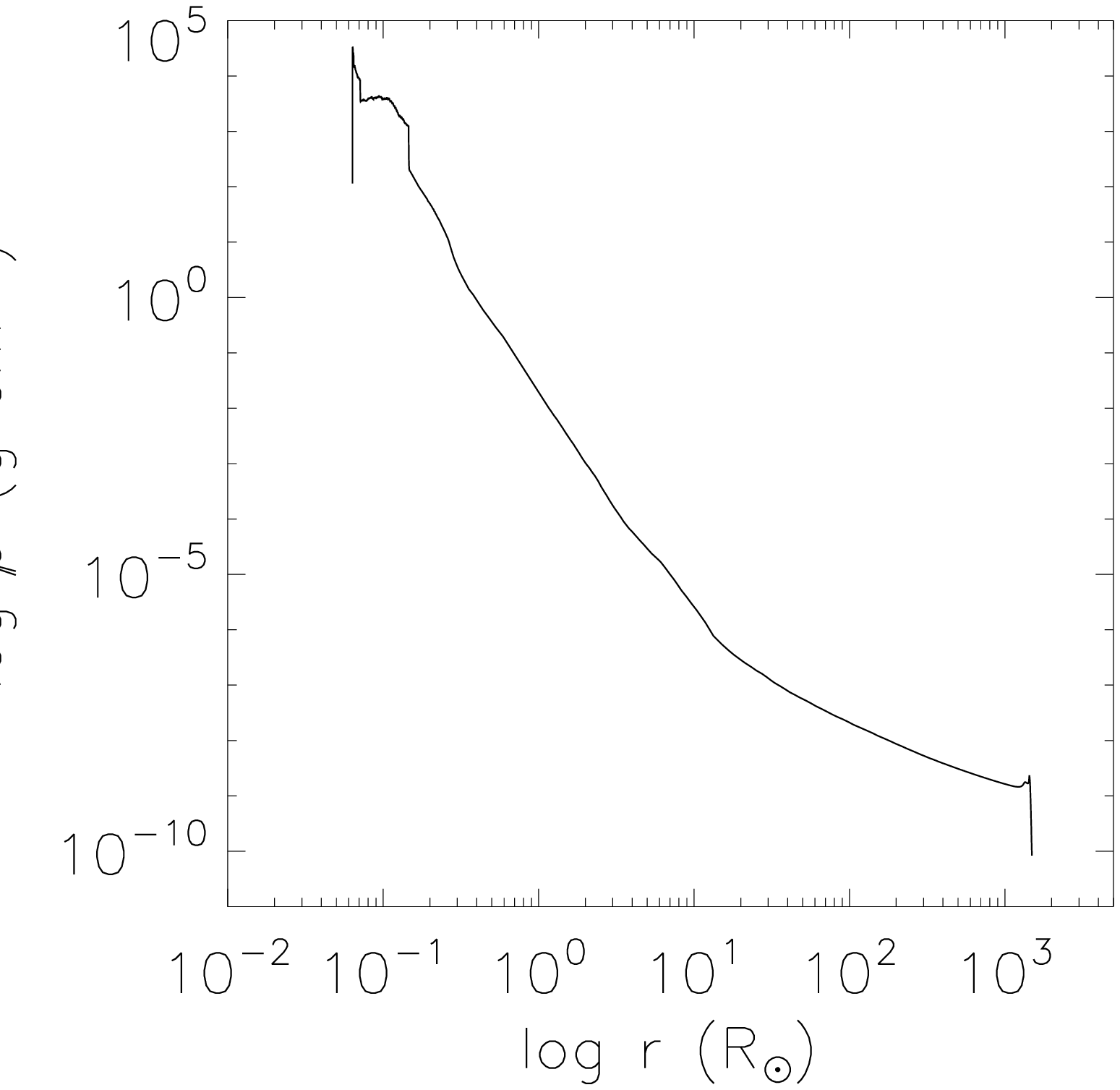}
\includegraphics[width=5.4cm]{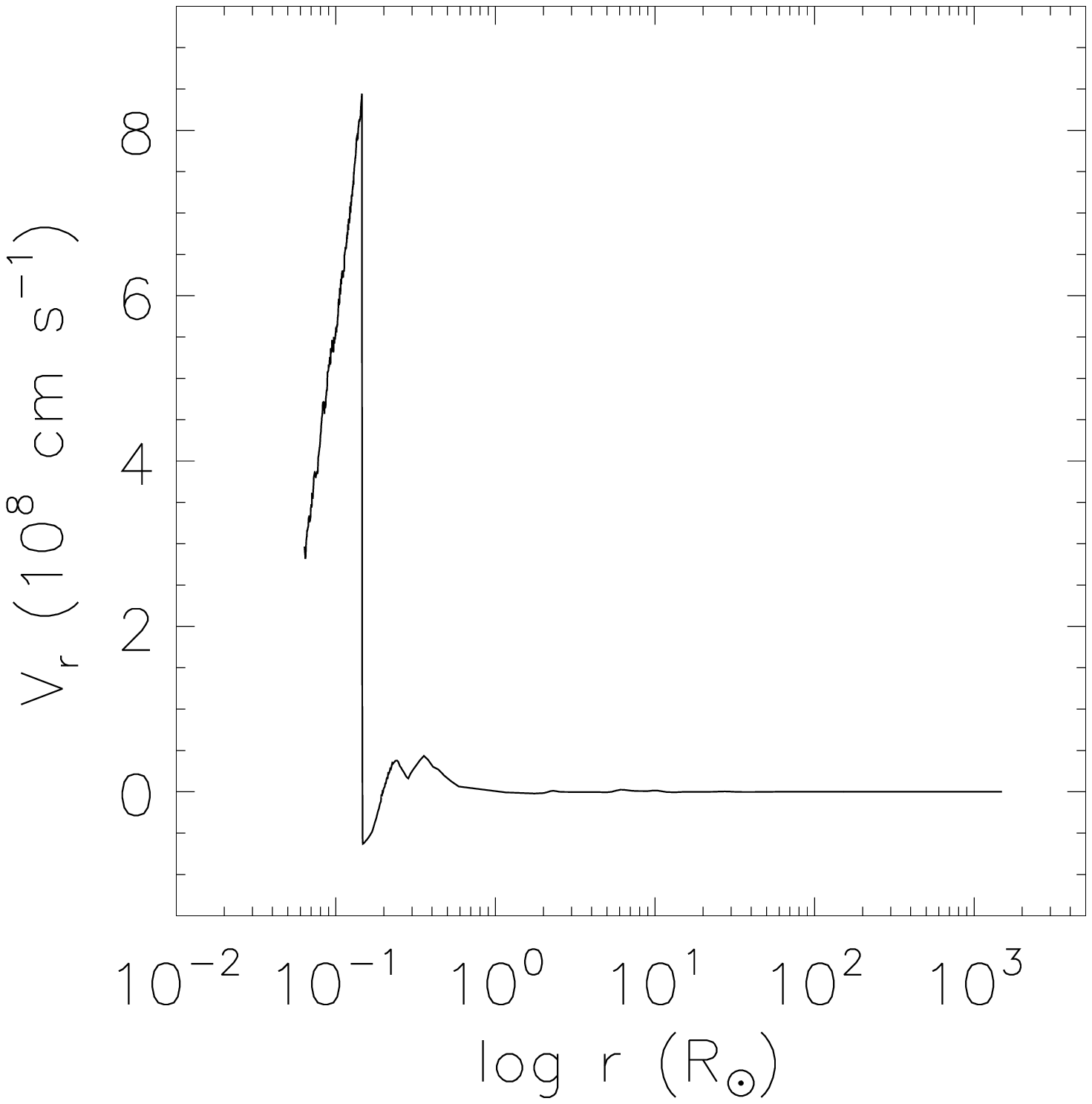}
\includegraphics[width=5.4cm]{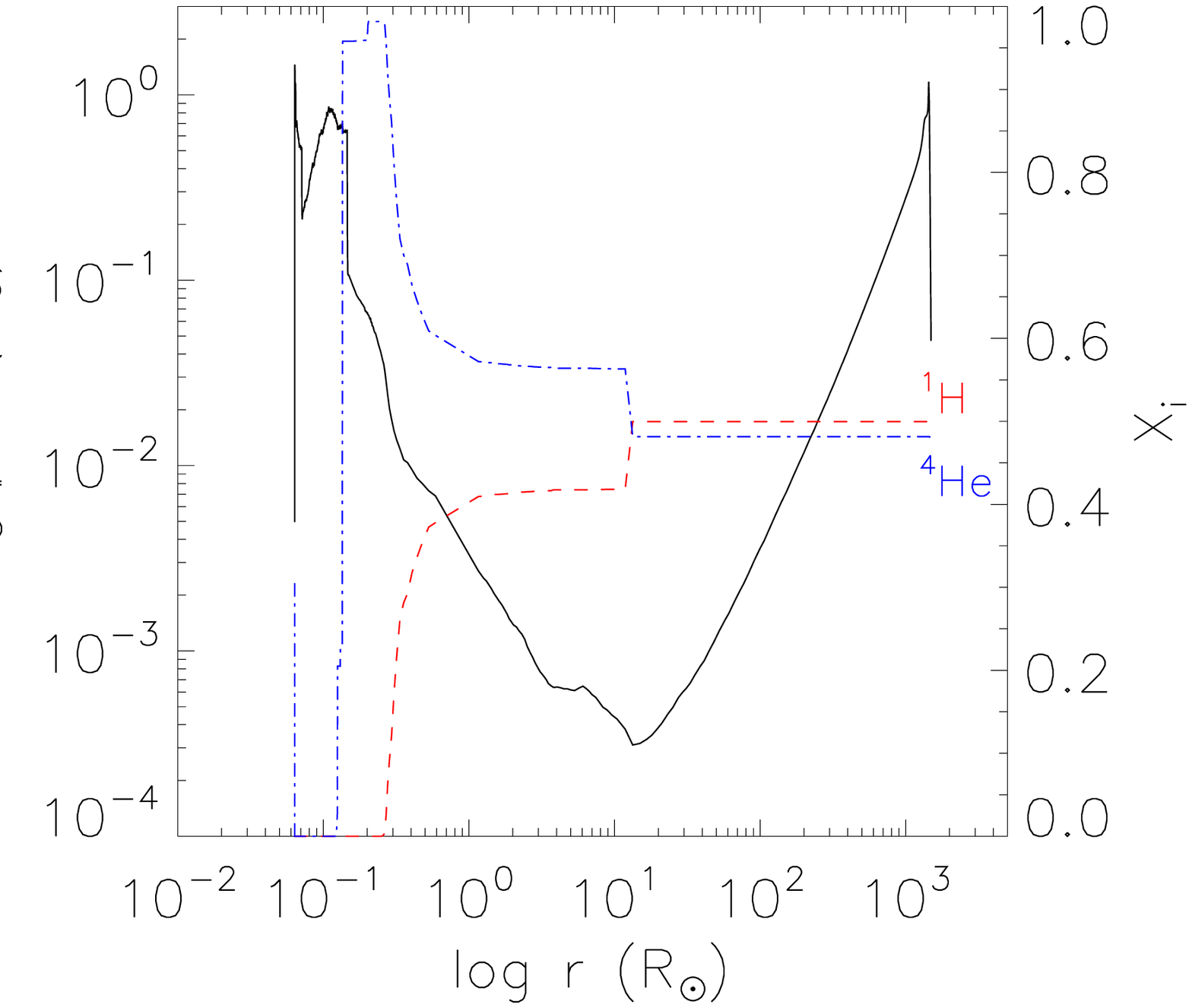}
\includegraphics[width=5.4cm]{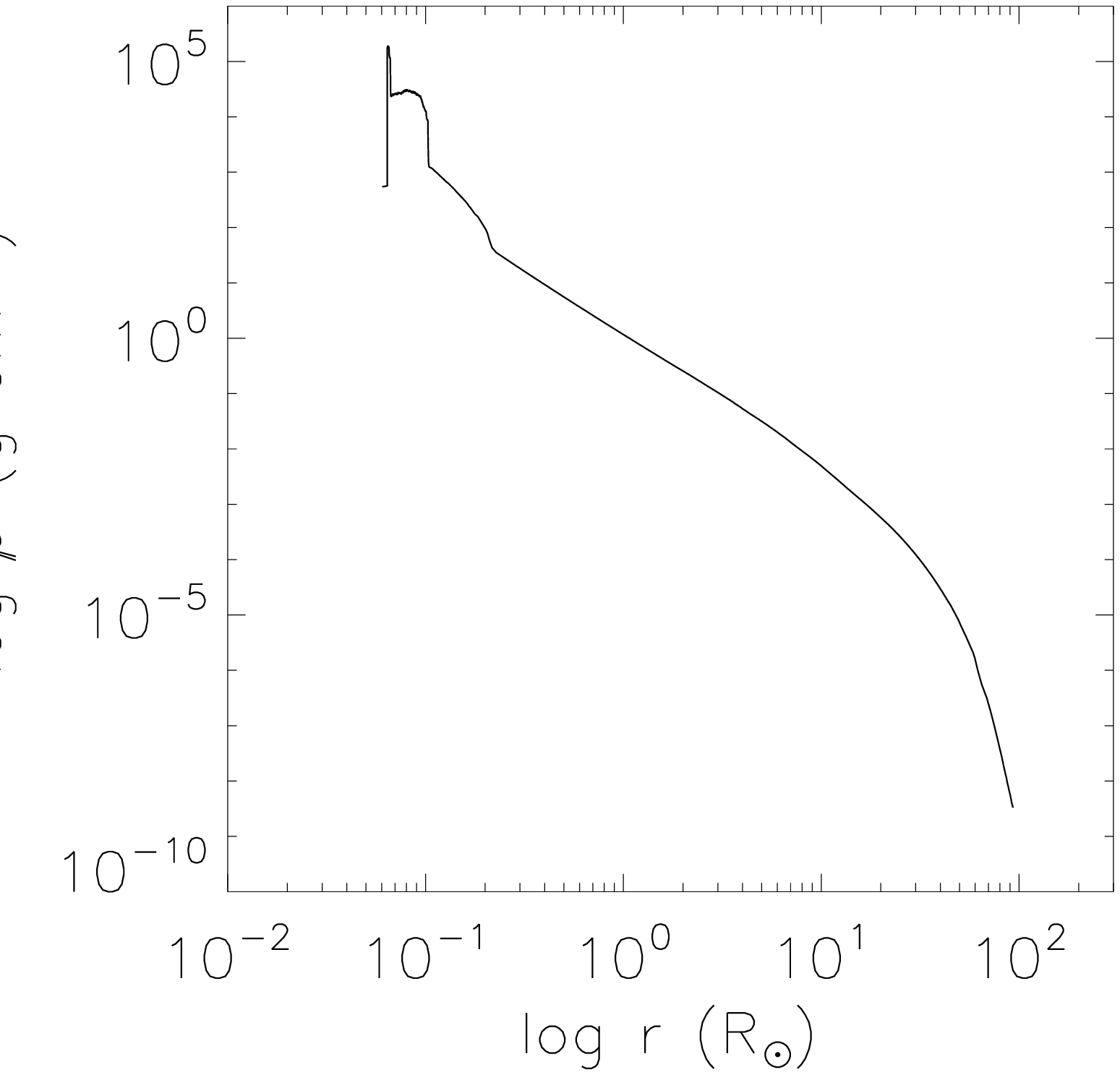}
\includegraphics[width=5.4cm]{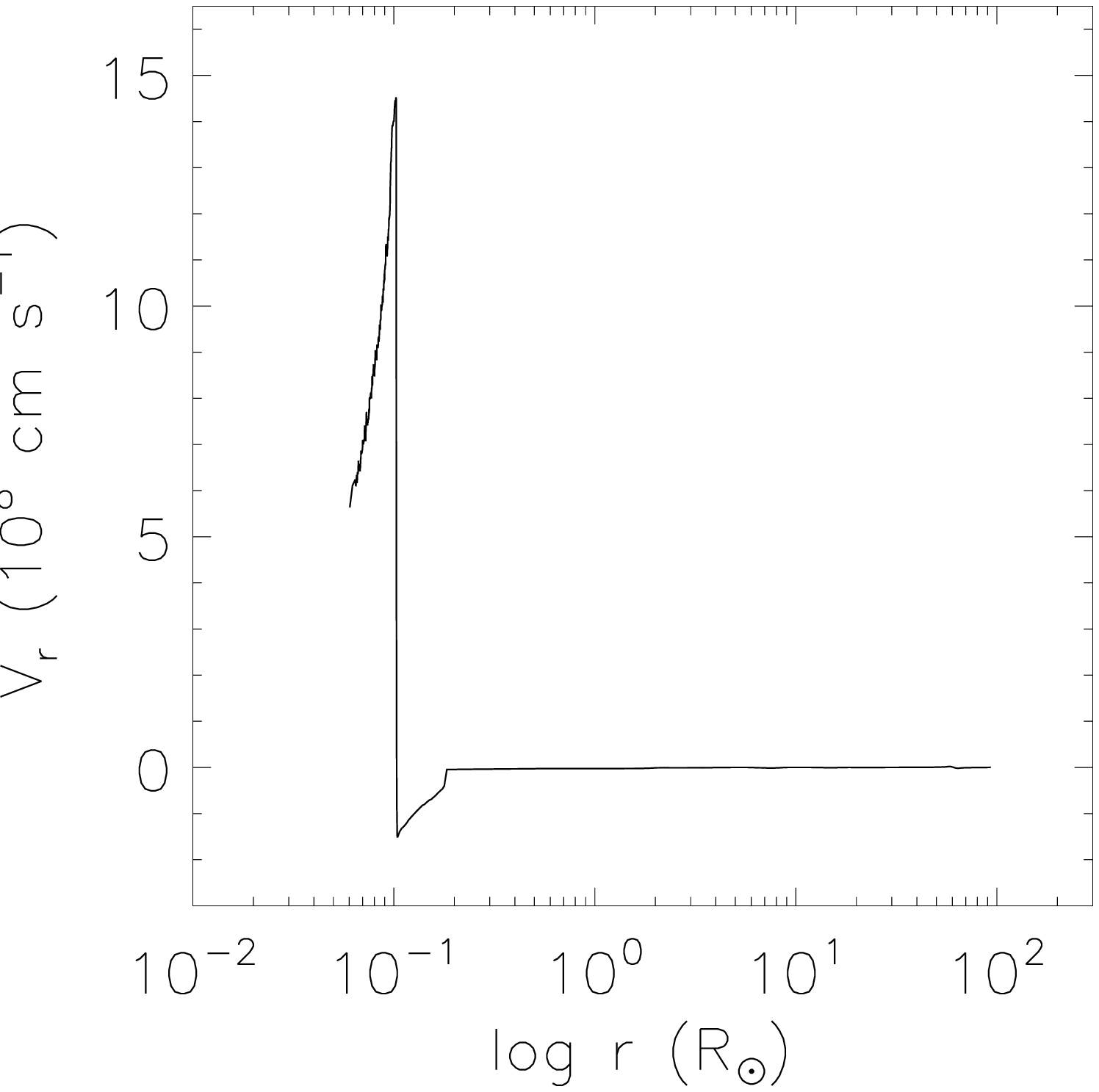}
\includegraphics[width=5.4cm]{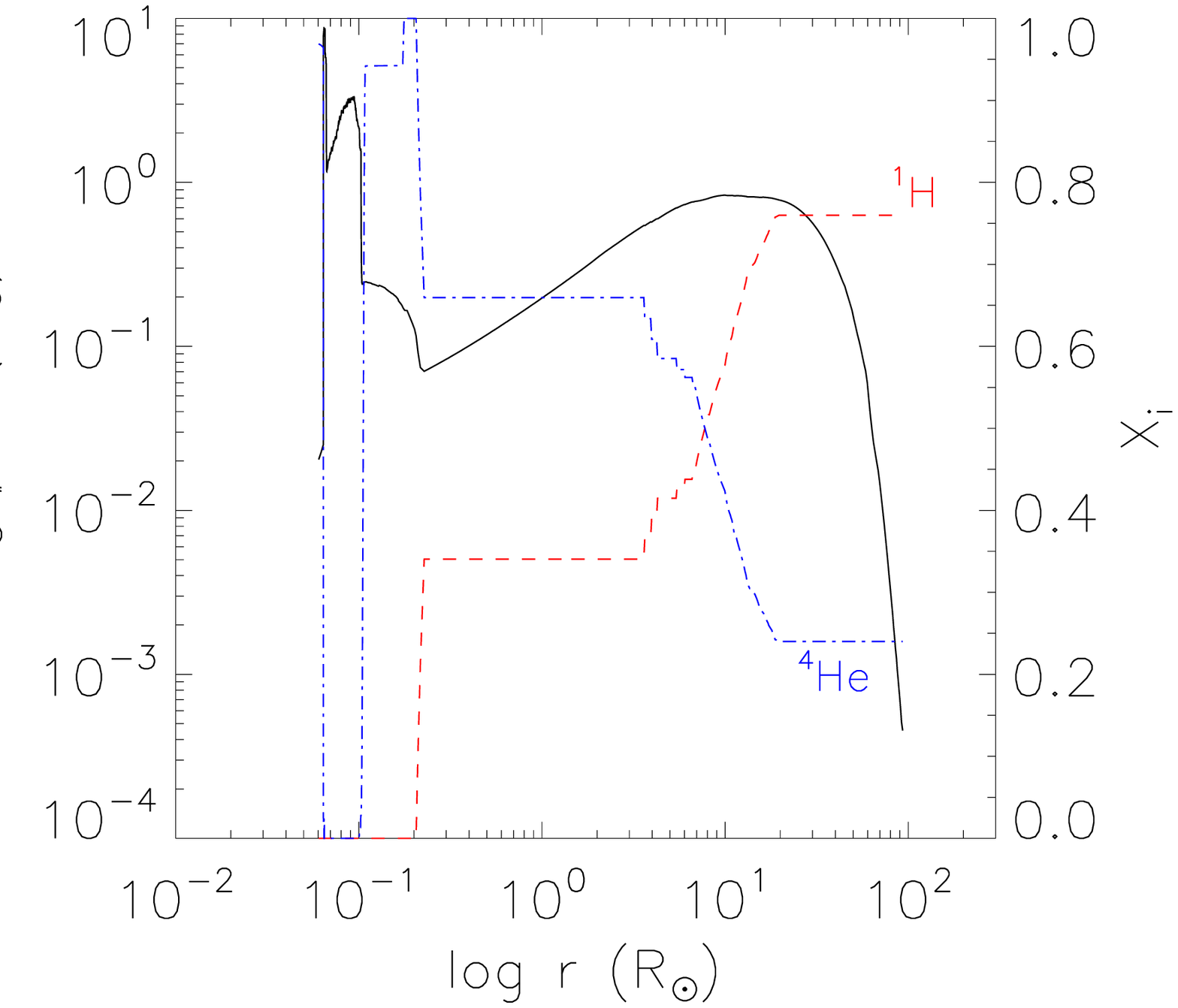}
\caption{
The internal structure of our progenitor S26 (top panels) and U50 (bottom panels) at the time of mapping the explosion simulation from one dimension (1D) to
three dimensions (3D).
The left and the central panels illustrate the density ($\rho$) and the radial velocity ($V_r$) profiles, respectively.
The right panels show the variation of the quantity $\rho r^3$ inside our progenitors, as well as the $^1$H and $^4$He abundances.}
\label{fig:progenitor_profiles}
\end{center}
\end{figure}

To model the core collapse and the launch of a forward shock, we use a 1D Lagrangian code to follow the
collapse of our progenitors through the core bounce.
The code was originally described in \cite{herant...94}.
It includes a broad set of equation of state to handle the wide range of densities in the collapse phase.
To model the neutrino transport, the code uses a gray flux-limited diffusion scheme \citep[see][for details]{herant...94, fryer99}
and considers the transport of three neutrino species: $\nu_e$, $\bar{\nu}_e$ and $\nu_x$.
Here, $\nu_x$ corresponds to the $\mu$ and $\tau$ neutrinos and their antiparticles, which are all treated equally.
The code also includes a nuclear network of 14 elements described in \cite{benz...89} to follow the energy generation.
Specifically, we model the collapse and the core bounce in 1D, then mimic a successful explosion by introducing an energy-deposition model.
The explosion is followed in 1D until the shock reaches roughly $10^{10}$\,cm. 
Our 1D energy-deposition model produces particular features in the fallback accretion rate that are different than multi-dimensional models.
As the energy in the injection region above the proto-neutron star increases, mass from the imploding star builds up above the energy injection region.
When this matter falls onto the injection region, it is shocked, increasing its density.
In a multi-dimensional simulation, some of this material will be convected through the energy-injection region and accreted onto the proto-neutron star.
In 1D, this material remains above the energy-injection region.
When the explosion is launched, this material is re-shocked by the supernova explosion.
This double-shocked material has higher density than the rest of the ejecta (see Figure\;\ref{fig:progenitor_profiles}).
As we shall see in our simulation results, this density spike causes a spike in the fallback rate.

\begin{deluxetable}{lcccccccc}
\tabletypesize{\tiny}
\tablewidth{16. cm}
\tablecolumns{9}
\tablecaption{Summary of Simulations}
\tablehead{\colhead{Run} & \colhead{Progenitor Model} &
\colhead{Resolution\tablenotemark{a}} & \colhead{$f$\tablenotemark{b}} & \colhead{$\Theta$ (deg)\tablenotemark{b}} &
\colhead{$\eta$\tablenotemark{c}} & \colhead{$E_{explosion}\,(10^{51} erg)$\tablenotemark{d}} &
\colhead{$M_{co,f} (M_\odot)$\tablenotemark{e}} & \colhead{$M_{fb} (M_\odot)$\tablenotemark{f}}
}
\startdata
\emph{s26\_symm (canonical)} & S26 & 1.2M & 1 & 0 & 1 & 2.78 & 3.00 & 1.02\\

s26\_f2th20 & S26 & 1.2M & 2 & 20 & 1 & 2.78 & 3.04 & 1.06\\
s26\_f2th40 & S26 & 1.2M & 2 & 40 & 1 & 2.78 & 3.15 & 1.17\\
s26\_f3th20 & S26 & 1.2M & 3 & 20 & 1 & 2.79 & 3.18 & 1.20\\
s26\_f3th40 & S26 & 1.2M & 3 & 40 & 1 & 2.79 & 3.67 & 1.69\\
s26\_f5th20 & S26 & 1.2M & 5 & 20 & 1 & 2.80 & 3.71 & 1.73\\
s26\_f5th40 & S26 & 1.2M & 5 & 40 & 1 & 2.80 & 4.69 & 2.71\\

s26\_symm\_n4 & S26 & 1.2M & 1 & 0 & 0.4 & 0.627 & 5.88 & 3.90\\
s26\_symm\_n5 & S26 & 1.2M & 1 & 0 & 0.5 & 0.945 & 4.74 & 2.76\\

s26\_f2th20\_n5 & S26 & 1.2M & 2 & 20 & 0.5 & 0.954 & 4.95 & 2.97\\
s26\_f2th40\_n5 & S26 & 1.2M & 2 & 40 & 0.5 & 0.974 & 5.28 & 3.30\\
s26\_f3th20\_n5 & S26 & 1.2M & 3 & 20 & 0.5 & 0.975 & 5.33 & 3.35\\
s26\_f3th40\_n5 & S26 & 1.2M & 3 & 40 & 0.5 & 1.02 & 5.85 & 3.87\\
s26\_f5th20\_n5 & S26 & 1.2M & 5 & 20 & 0.5 & 1.04 & 6.09 & 4.11\\
s26\_f5th40\_n4 & S26 & 1.2M & 5 & 40 & 0.4 & 0.782 & 6.71 & 4.73\\
s26\_f5th40\_n5 & S26 & 1.2M & 5 & 40 & 0.5 & 1.09 & 6.30 & 4.32\\

s26\_symm\_p10M & S26 & 9.5M & 1 & 0 & 1 & 2.75 & 3.10 & 1.12\\
s26\_f5th40\_n5\_p10M & S26 & 9.5M & 5 & 40 & 0.5 & 1.07 & 6.40 & 4.42\\

\tableline

\emph{u50\_symm (canonical)} & U50 & 1.2M & 1 & 0 & 1 & 20.3 & 6.69 & 4.43\\

u50\_f2th20 & U50 & 1.2M & 2 & 20 & 1 & 20.3 & 6.67 & 4.41\\
u50\_f2th40 & U50 & 1.2M & 2 & 40 & 1 & 20.4 & 6.52 & 4.26\\
u50\_f3th20 & U50 & 1.2M & 3 & 20 & 1 & 20.4 & 6.88 & 4.62\\
u50\_f3th40 & U50 & 1.2M & 3 & 40 & 1 & 20.4 & 7.07 & 4.81\\
u50\_f5th20 & U50 & 1.2M & 5 & 20 & 1 & 20.4 & 7.51 & 5.25\\
u50\_f5th40 & U50 & 1.2M & 5 & 40 & 1 & 20.1 & 8.29 & 6.03\\

u50\_symm\_n4 & U50 & 1.2M & 1 & 0 & 0.4 & 3.79 & 13.0 & 10.7\\
u50\_symm\_n5 & U50 & 1.2M & 1 & 0 & 0.5 & 6.33 & 10.0 & 7.74\\

u50\_f2th40\_n5 & U50 & 1.2M & 2 & 40 & 0.5 & 6.36 & 11.0 & 8.74\\
u50\_f3th20\_n5 & U50 & 1.2M & 3 & 20 & 0.5 & 6.43 & 11.1 & 8.84\\
u50\_f3th40\_n5 & U50 & 1.2M & 3 & 40 & 0.5 & 6.50 & 12.2 & 9.94\\
u50\_f5th20\_n5 & U50 & 1.2M & 5 & 20 & 0.5 & 6.64 & 13.2 & 10.9\\
u50\_f5th40\_n4 & U50 & 1.2M & 5 & 40 & 0.4 & 4.54 & 18.2 & 15.9\\
u50\_f5th40\_n5 & U50 & 1.2M & 5 & 40 & 0.5 & 6.75 & 15.8 & 13.5\\
\enddata
\tablenotetext{a}{number of SPH particles (in million) used in the simulation}
\tablenotetext{b}{parameters describing the induced asymmtries (see \S\,\ref{sec:simulatio_runs})}
\tablenotetext{c}{kinetic and internal energy scaling factor for the material behind the forward shock at the time of 3D mapping (see \S\,\ref{sec:simulatio_runs})}
\tablenotetext{d}{explosion energy}
\tablenotetext{e}{final mass of the central compact object (at $\sim 0.5$\;yr after the launch of the explosion)}
\tablenotetext{f}{total mass of fallback}
\label{tab:simulations_summary}
\end{deluxetable}

\subsection{Three Dimensional Mapping}

Less than 10\,s after the launch of a forward shock, we map the explosion of our progenitors into 3D and follow the rest of the
explosion in 3D.
Specifically, this mapping is performed when the forward shock reaches $\sim 10^{10}$\;cm.
For the progenitor S26, this means the forward shock has propagated into the He-rich layer.
For the progenitor U50, we implement this mapping when the forward shock hits the O/He interface.
Figure~\ref{fig:progenitor_profiles} illustrates the internal structure of our progenitors at the time of 3D mapping.

To model the rest of the explosion in 3D, we use the 3D Lagrangian hydrodynamic code SNSPH \citep{fryer...06}.
This code uses a particle-based algorithm and adopts Benz version \citep{benz84, benz88, benz90} of smooth particle hydrodynamics (SPH)
to model the Euler Equations.
SNSPH is built on the parallel tree code developed by \cite{warren&salmon93, warren&salmon95}.
SNSPH also includes a network of 20 elements described in \cite{ellinger...12} to follow the change in the chemical composition and
the specific internal energy of each SPH particle due to nuclear burning or radioactive decay.
In addition, we adopt the scheme of weighted Voronoi tesselations \citep{diehl...12} to set up the initial SPH-related conditions for our 3D
explosion simulations.

\subsection{Inner Boundary Condition}
\label{sec:inner_bndry}

In our 3D explosion simulations, the central compact object located at the center is cut out and replaced by a spherical, absorbing inner boundary.
The gravitational effect of this compact object is mimicked with a central gravitational force term.
Replacing the central compact object with an inner boundary relaxes the Courant constraint on the time step and reduces the numerical
noise at the core of the exploding star.

Throughout our 3D explosion simulation, any SPH particles falling across the inner boundary are removed from the simulation.
In each time-step iteration, the masses of these removed SPH particles are added to the mass of the central compact object.
This absorptive boundary condition assumes that the matter flowing across the boundary quickly cools and compresses onto the proto-neutron star.
The cooling timescale for the high fallback rates in our calculations is typically fast with respect to the accretion timescale, validating our
inner boundary condition.
However, this accretion can drive outflows, providing pressure that we do not include in our calculations
\citep{fryeretal06, fryer09, lindner...12, milosavljevic...12, dexter13}.
Typically, these outflows will decrease our accretion rate by 10--30\%.
Hence, they will not change the accretion rate significantly for most of our models.
We will discuss the implications of these outflows on the nucleosynthetic yields and explosion luminosity in our discussion section
(\S\,\ref{sec:discussion}).

Furthermore, if the distance ($r_p$) between an SPH particle and the center of the inner boundary is less than twice of the smoothing
length ($h$) of that particle, the mass ($M_p$) of that particle will be accreted onto the central compact object at a rate given by
\begin{equation}
\dot{M}_p = \frac{M_p}{t_{ff}}.
\end{equation}
Here, $t_{ff}$ is the free-fall timescale and can be written as
\begin{equation}
t_{ff} = \sqrt{\frac{2 r^3_p}{G M_{CO}}},
\end{equation}
where $G$ is the gravitational constant and $M_{CO}$ is the mass of the central compact object.
\cite{zhang&fryer01} were the first to consider this accretion treatment when studying the merger of a He star and a black hole as a progenitor of
gamma ray burst.

If no SPH particle falls across the inner boundary, the radius of the inner boundary is increased to 80\% of the distance
from its center to the nearest SPH particle.
In addition, the inner boundary is not allowed to shrink.

\begin{figure}
\epsscale{1.0}
\begin{center}
\plottwo{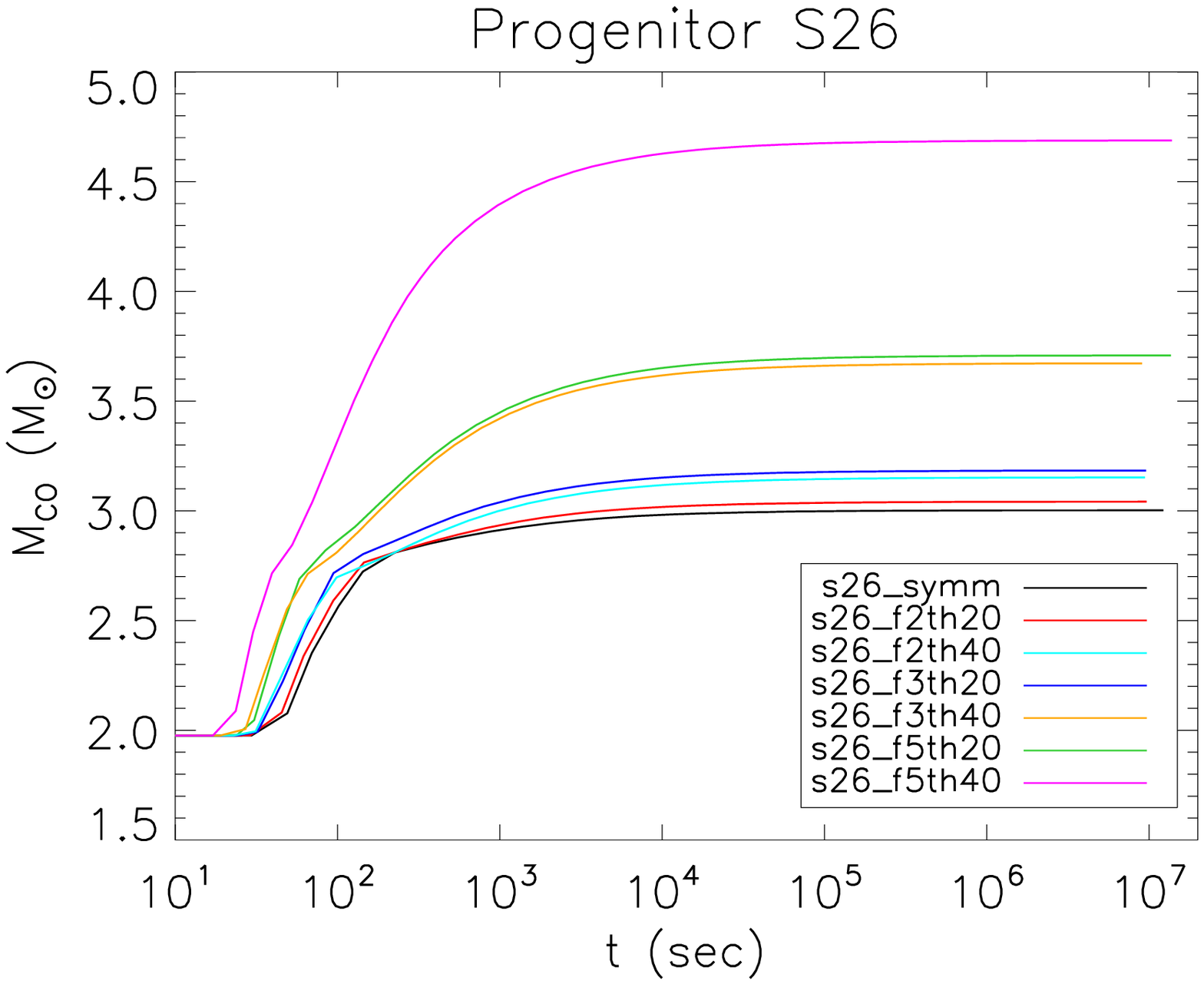}{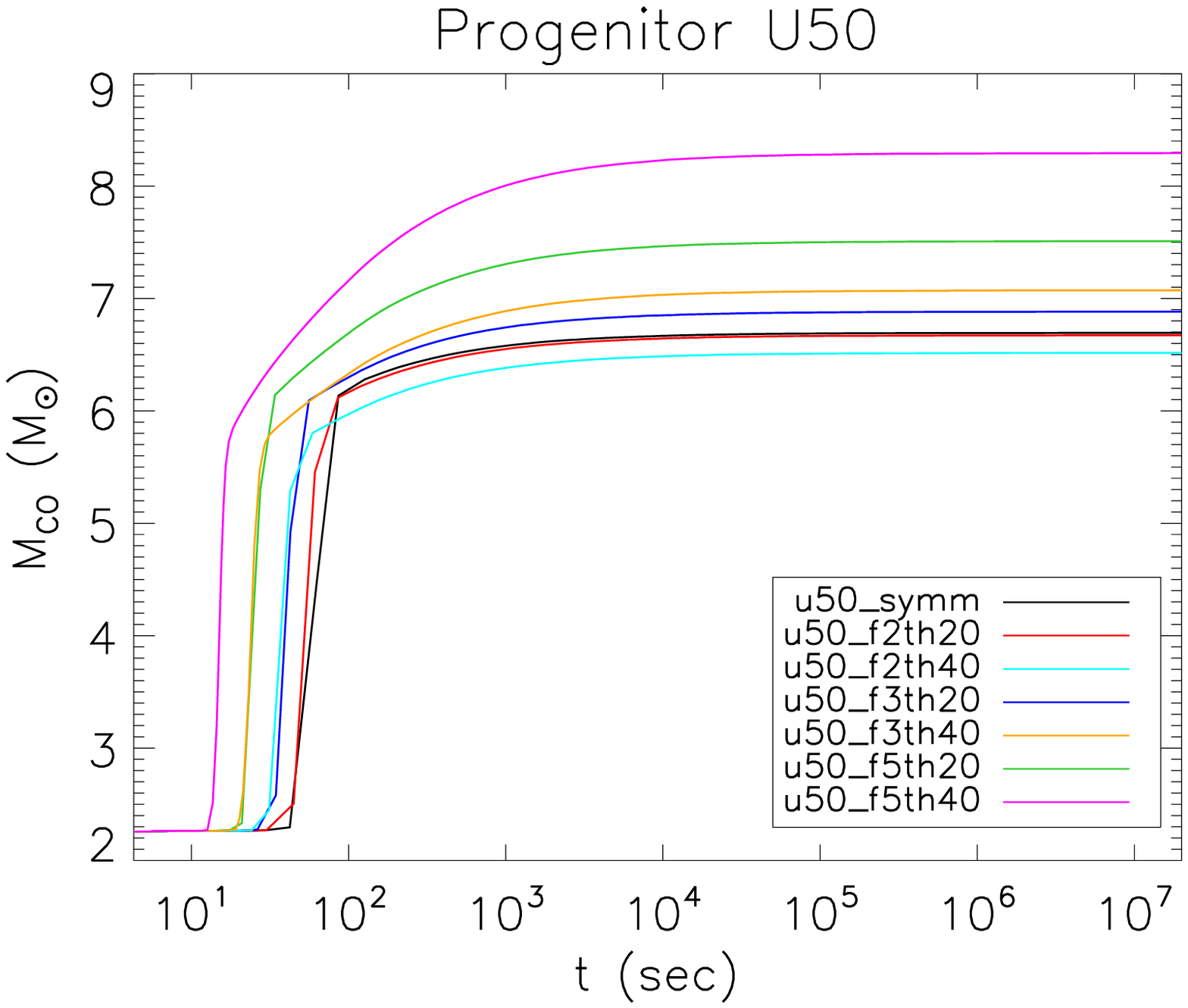}
\plottwo{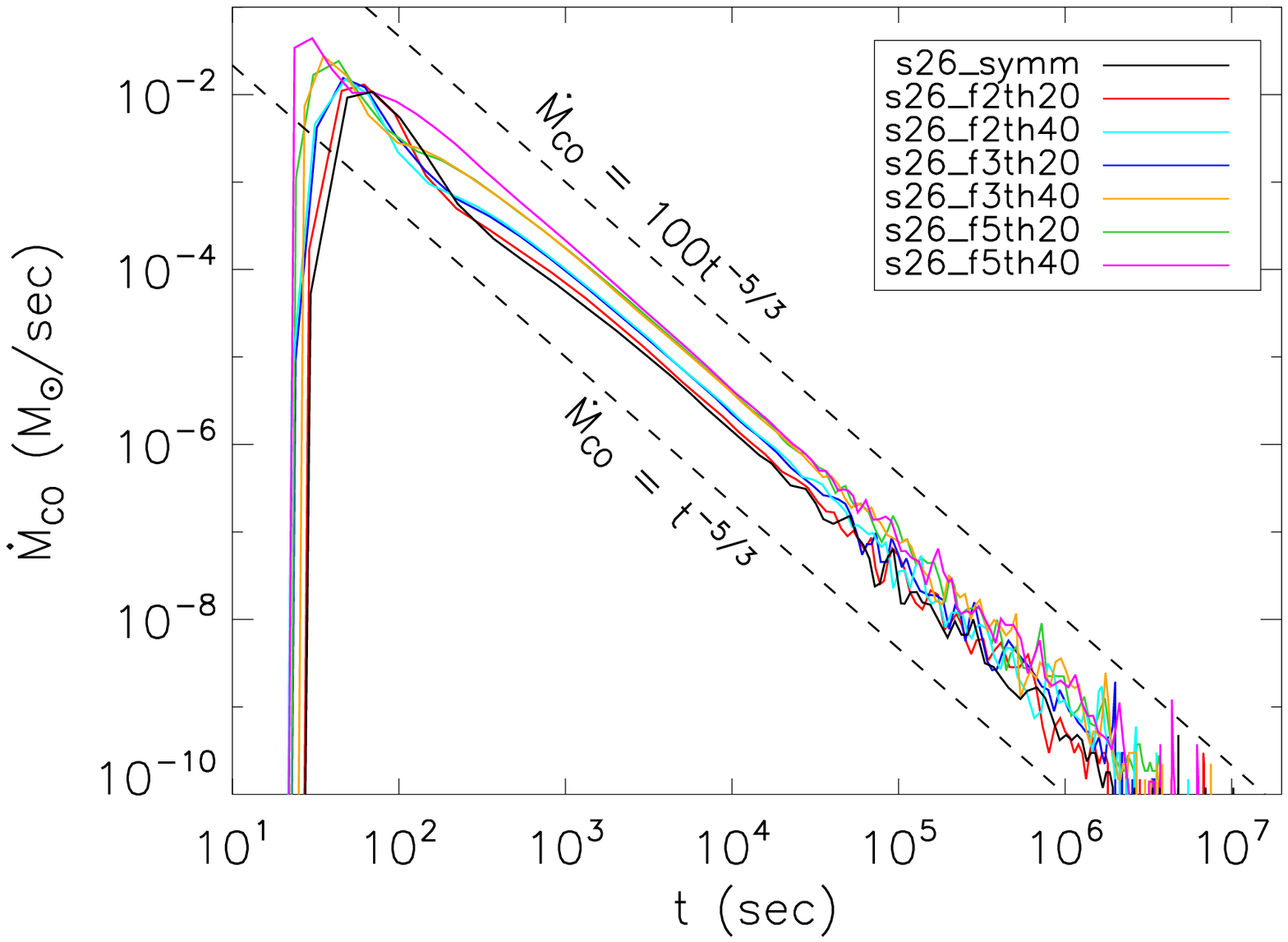}{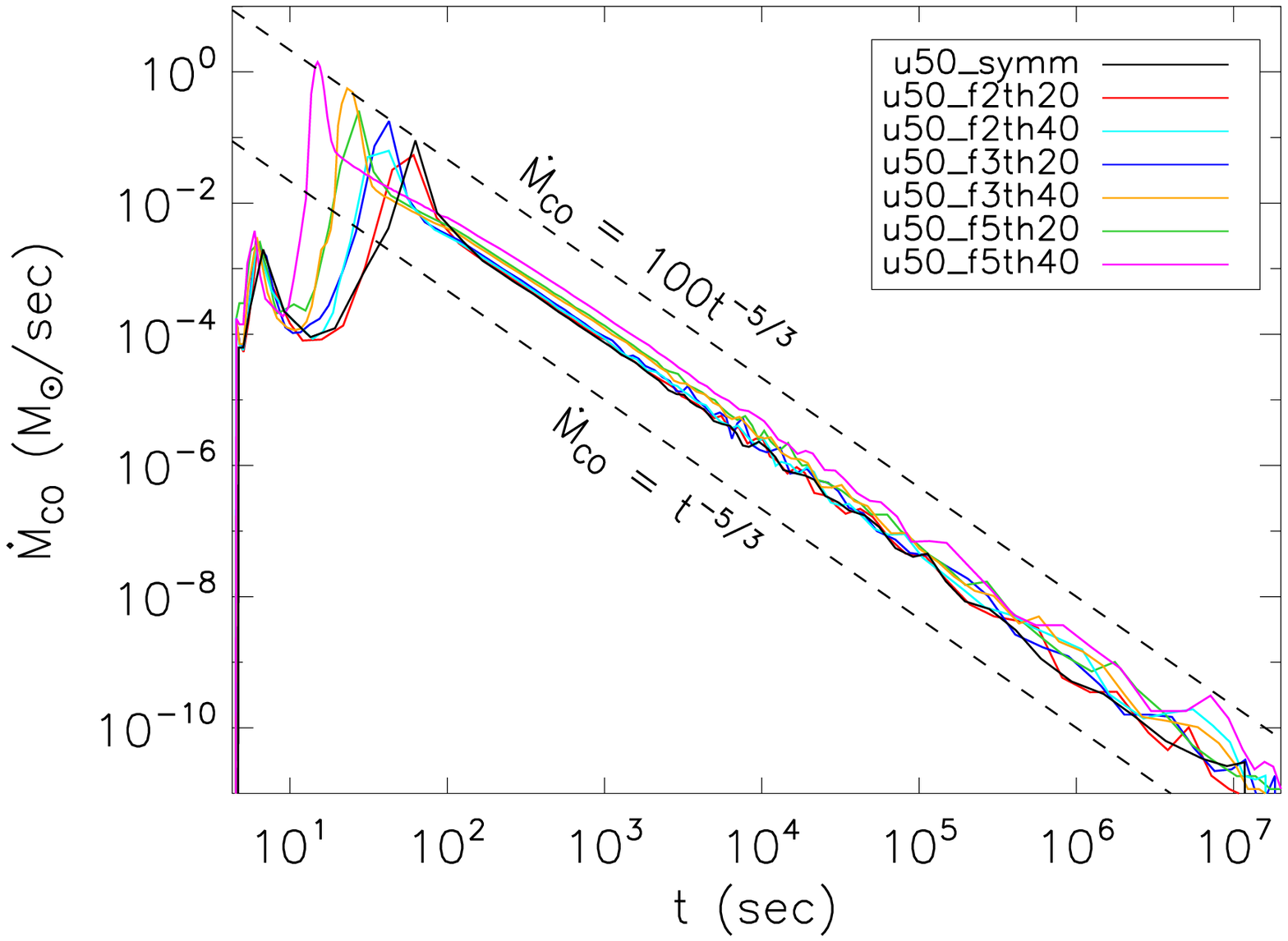}
\caption{The mass ($M_{co}$, top panels) and accretion rate ($\dot{M}_{co}$, bottom panels) of the central compact object vs. time (t)
elapsed since the start of the explosion for simulations with standard explosion energies ($\eta = 1$).
In all of these simulations, the rate of fallback ($\equiv \dot{M}_{co}$) always decays as $t^{-5/3}$ after $t \approx 100$\,s.
This indicates that fallback occurs primarily due to prompt fallback mechanism.} 
\label{fig:mco_plot1}
\end{center}
\end{figure}

\begin{figure}
\epsscale{0.5}
\begin{center}
\plotone{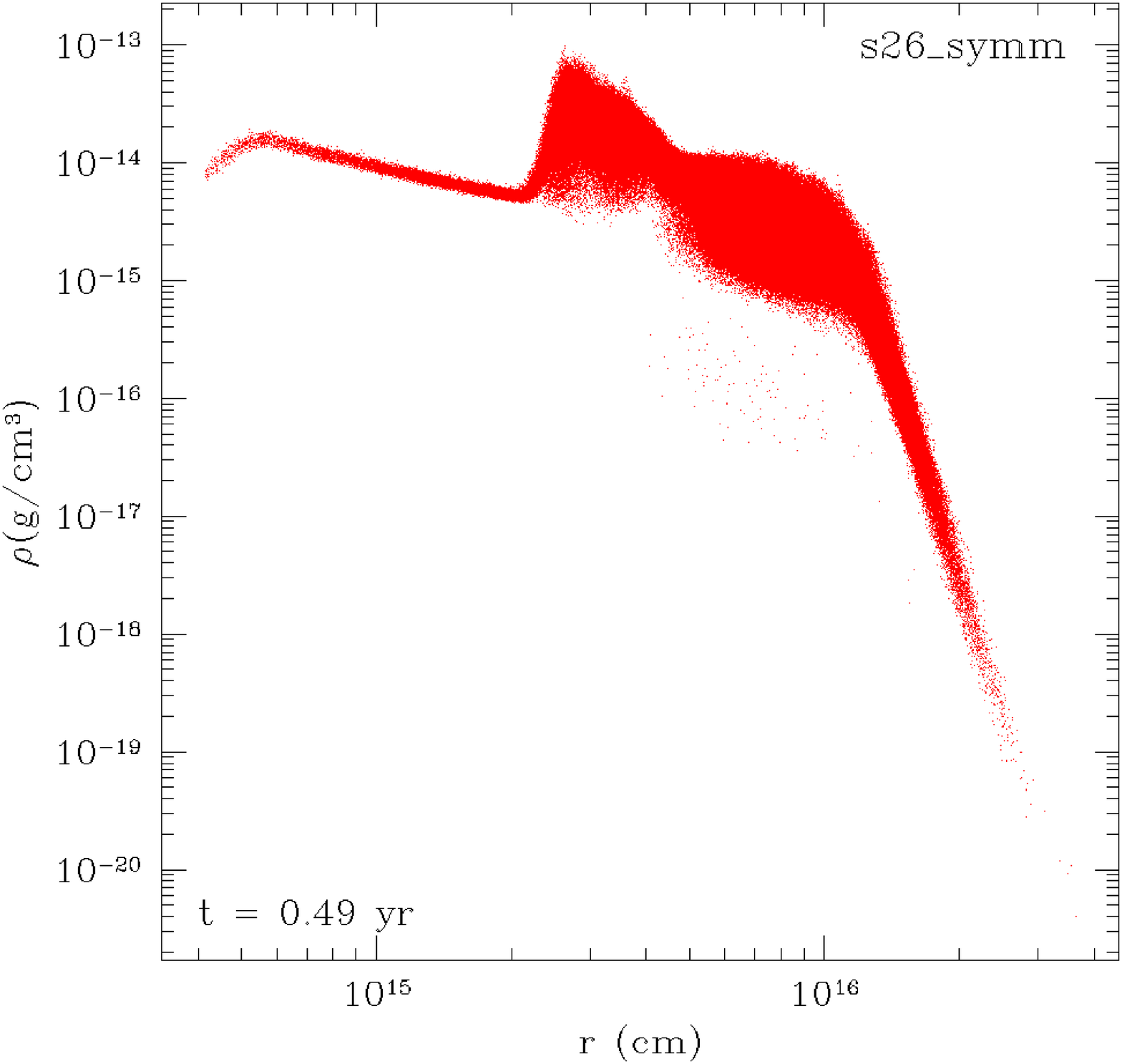}
\plotone{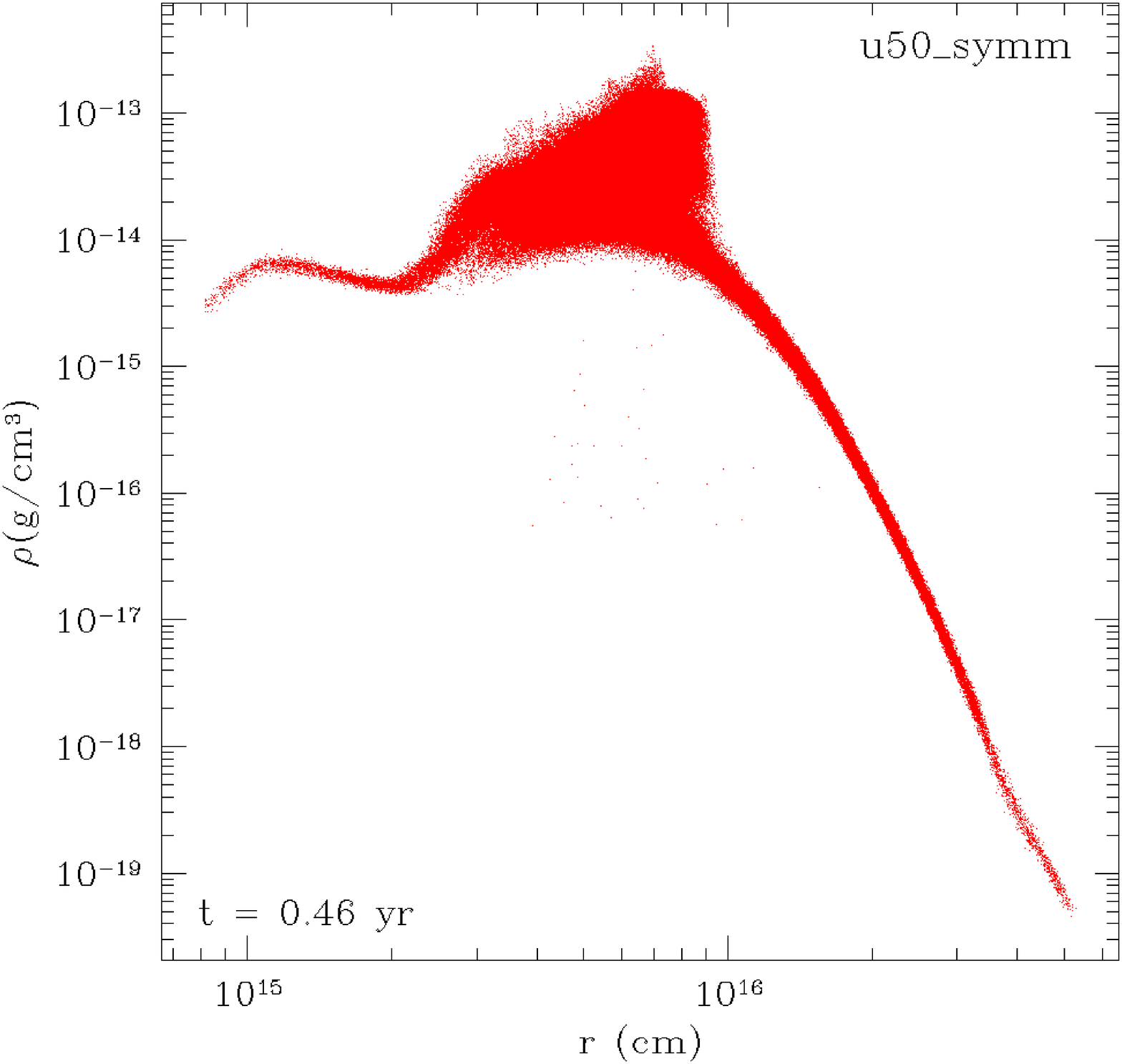}
\caption{
The density profiles of our canonical runs for progenitor S26 and U50 at the time when we terminate the simulations.
The inner boundaries are located at $4.1 \times 10^{14}$ and $8.2 \times 10^{14}$\;cm for the simulation s26\_symm and u50\_symm, respectively.
It is clear that the reverse shock in both simulations do not get close to the center (or the inner boundary).
Hence, the reverse shocks cannot produce any fallback.
}
\label{fig:symm_lrho}
\end{center}
\end{figure}

\subsection{Simulation Runs}
\label{sec:simulatio_runs}

Unless specified, all of our 3D explosion simulations use roughly 1.2 million SPH particles.
To account for the range in physical densities in our exploding progenitor stars (see Figure\;\ref{fig:progenitor_profiles}),
SPH particle masses at the beginning of our 3D explosion simulations range from 3.0 to $24 \times 10^{-6} M_\odot$ for the progenitor S26,
and from 0.15 to $92 \times 10^{-6} M_\odot$ for the progenitor U50.
The average SPH particle mass in the initial 3D configuration for progenitor S26 and U50 are 9.1 and $39 \times 10^{-6} M_\odot$, respectively.
This choice is justified in \S\,\ref{subsec:convergence_test}.
Our canonical runs are the symmetric explosions of the progenitor S26 and U50.

From the current understanding of core-collapse supernova, the explosion itself could be asymmetric \citep[see e.g.][]{janka12}.
Hence, we are also interested in studying the nature of fallback with asymmetric blast waves.
We follow \cite{hungerford...05} and impose single-lobe asymmetries in several simulations of each progenitor.
Specifically, we artificially alter the velocities of the SPH particles behind the forward shock at the time of 3D mapping.
Our input asymmetries are created using a conical geometry, which can be described by two parameters:
(1) $\Theta~$=~the opening angle of an enhanced explosion cone and
(2) $f$~=~the ratio of the in-cone velocity to the out-of-cone velocity.
By requiring that the total kinetic energy of the exploding star remains unchanged, the in-cone and out-of-cone radial velocities are set to be
\begin{equation}
V_{in-cone} = f \left [ \frac{1-f^2}{2} \cos (\Theta) + \frac{1+f^2}{2} \right ] ^{-1/2} V_{symm}
\end{equation}
\begin{equation}
V_{out-of-cone} = \left [ \frac{1-f^2}{2} \cos (\Theta) + \frac{1+f^2}{2} \right ] ^{-1/2} V_{symm}\;.
\end{equation}
Here, $V_{symm}$ is the radial velocity given by our 1D explosion models at the time of 3D mapping.
We consider the same values of $f$ and $\Theta$ as \cite{hungerford...05}, which are guided by the multidimensional core-collapse simulations
of \cite{blondin...03} and \cite{scheck...04}.

Finally, we are also interested in how the fallback depends on the explosion energy.
Instead of re-modeling the launch of an explosion in 1D using a different duration or efficiency of energy deposition, we artificially alter the explosion
energy by scaling the internal and kinetic energy of all SPH particles behind the forward shock by a constant ($\eta$) at the time of 3D mapping.
The internal energy of each SPH particle includes the thermal and the radiation energy.
Artificially altering the explosion energy has the advantage that it will not change the internal structure of the exploding progenitor star, except for the
radial velocity and the temperature profiles behind the forward shock.

Our suite of simulations is listed in Table~\ref{tab:simulations_summary}.
The stars in our simulations are all exploded into a vacuum and we do not include any circumstellar or interstellar medium material.
We terminate our explosions 0.5\;yr after the launch of the explosion and, in this time, we expect less than $10^{-6}$--$10^{-4} M_\odot$ to be swept up
(even for our massive stars).
Objects like SN2010jl, which have considerable surrounding material, are rare cases \citep{ofek...13}.
Hence, for most of our calculations, the vacuum assumption is valid.

\section{Results}
\label{sec:result}

\subsection{Canonical Runs}

Our canonical 3D explosion simulation (s26\_symm) of the progenitor S26 starts at $t = 9\,s$, where $t$ is the time since the launch of the explosion.
Because of the strong shock related to the density spike due to our 1D energy deposition model, there is no fallback until $t = 29\,s$.
This late initial fallback is an artifact of our explosion.
In driving a strong explosion, we have deposited considerable energy in all ejecta.
In a multi-dimensional simulation, some material would have less energy and would fall back earlier, providing a more continuous fallback distribution
from the onset of the explosion to late times.
The fallback in our explosion model peaks at $t = 53$\,s when the matter in the density spike falls across the inner boundary, and begins to decay at a
rate proportional to $t^{-5/3}$ (see Figure\,\ref{fig:mco_plot1}).
Meanwhile, the forward shock continues to accelerate as it propagates outwards in the He-rich layer of the stellar envelope.
At $t \approx 370$\,s, the forward shock reaches its maximum speed of roughly 31000\,km/s at the He/H interface and has a radius of $10^{12}$\;cm.
Meanwhile, 0.87\;$M_\odot$ of the ejecta has fallen back onto the proto-neutron star.
Consequently, there are 1.06 million SPH particles left in the simulation, of which 0.63 million particles are behind the forward shock.
Then, the forward shock starts to decelerate, as it propagates into the region where the quantity $\rho r^3$ rises exponentially
(see Figure\,\ref{fig:progenitor_profiles}).
At the same time, a reverse shock is reflected towards the center.
However, this reverse shock freezes in the expanding ejecta before ever propagating close to the center (see Figure~\ref{fig:symm_lrho}).
Hence, the reverse shock does not have any influence on the amount of fallback. 
As shown in Figure~\ref{fig:mco_plot1}, the central compact object has accreted most of the fallback material in the first 1000\,s after the launch of
the explosion.

For our canonical run of the progenitor U50 (u50\_symm), the 3D simulation starts at $t = 4\,s$.
Unlike the canonical run of our progenitor S26, fallback occurs immediately after the start of our 3D simulation, as the material behind the density
spike resulted from our 1D energy deposition model falls across the inner boundary.
The fallback of this material produces a small peak in the fallback rate at $t = 7\,s$ (see Figure\,\ref{fig:mco_plot1}).
At $t = 62$\,s, the fallback reaches its highest peak when the material in the density spike falls across the inner boundary.
Then, the fallback begins to decay as $t^{-5/3}$.
On the other hand, the forward shock continues to accelerate when it is still traveling in the He-rich layer.
It acquires its maximum speed of 21000\,km/s when it hits the convective H-burning shell at $t = 9\,s$, before the fallback peaks.
At the same time, the forward shock reaches $1.6 \times 10^{10}$\;cm.
Since $< 0.01\;M_\odot$ of the ejecta has fallen back onto the proto-neutron star, there are still roughly 1.2 million SPH particles less
in the simulation, of which 0.44 million particles are behind the forward shock.
Then, the forward shock begins to decelerate, as the quantity $\rho r^3$ increases rapidly with $r$ (see Figure\,\ref{fig:progenitor_profiles}).
Due to this deceleration, a reverse shock is driven towards the center. 
Compared to the canonical run of the progenitor S26, this reverse shock appears earlier and well before the fallback reaches its
highest peak.
However, this reverse shock also fails to get close to the center before it freezes in the expanding ejecta (see Figure~\ref{fig:symm_lrho}).
As with our S26 model, the reverse shock again has no influence on the amount of fallback material.
Figure~\ref{fig:mco_plot1} illustrates that most of the fallback material is accreted onto the central compact object in the first 100\,s after the launch of
the explosion.

As a result, the fallback in our canonical runs of both progenitors are solely due to the prompt fallback mechanism.

\subsection{Explosion Asymmetries}

In our simulations with asymmetric blast waves (s26\_f`$A$'th`$B$' and u50\_f`$A$'th`$B$', where `$A$' can be 2, 3 or 5 and `$B$' can be 20 or 40,
see Table\,\ref{tab:simulations_summary}), we find that the primary effect of asymmetries is to change the amount of fallback, but the timing and
evolution of fallback is very similar to our symmetric models (see Figure\,\ref{fig:mco_plot1}).
Although the induced explosion asymmetries help the reverse shock propagate closer to the center, it is still not close enough for the
reverse shock to have an influence on fallback. 
In these simulations, the amount of fallback material is larger only because the velocity of the post-shock material is
altered in order to impart single-lobe explosion asymmetries.
By redistributing the kinetic energy to create a cone of enhanced explosion, the velocities of some out-of-cone material are decreased below the
local escape velocity, causing them to eventually fall back onto the central compact object.

\subsection{Less Energetic Explosions}

\begin{figure}
\epsscale{1.0}
\begin{center}
\plottwo{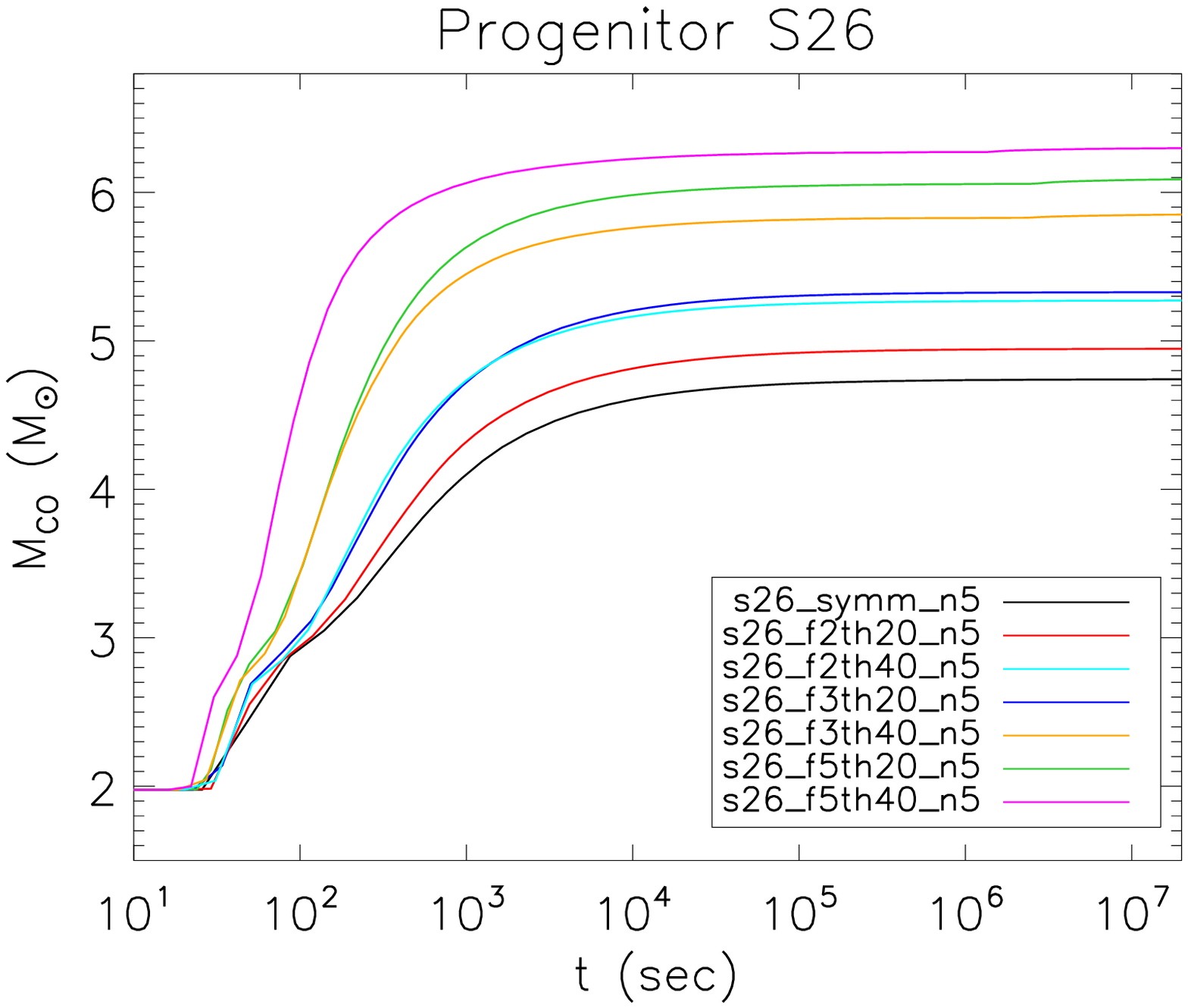}{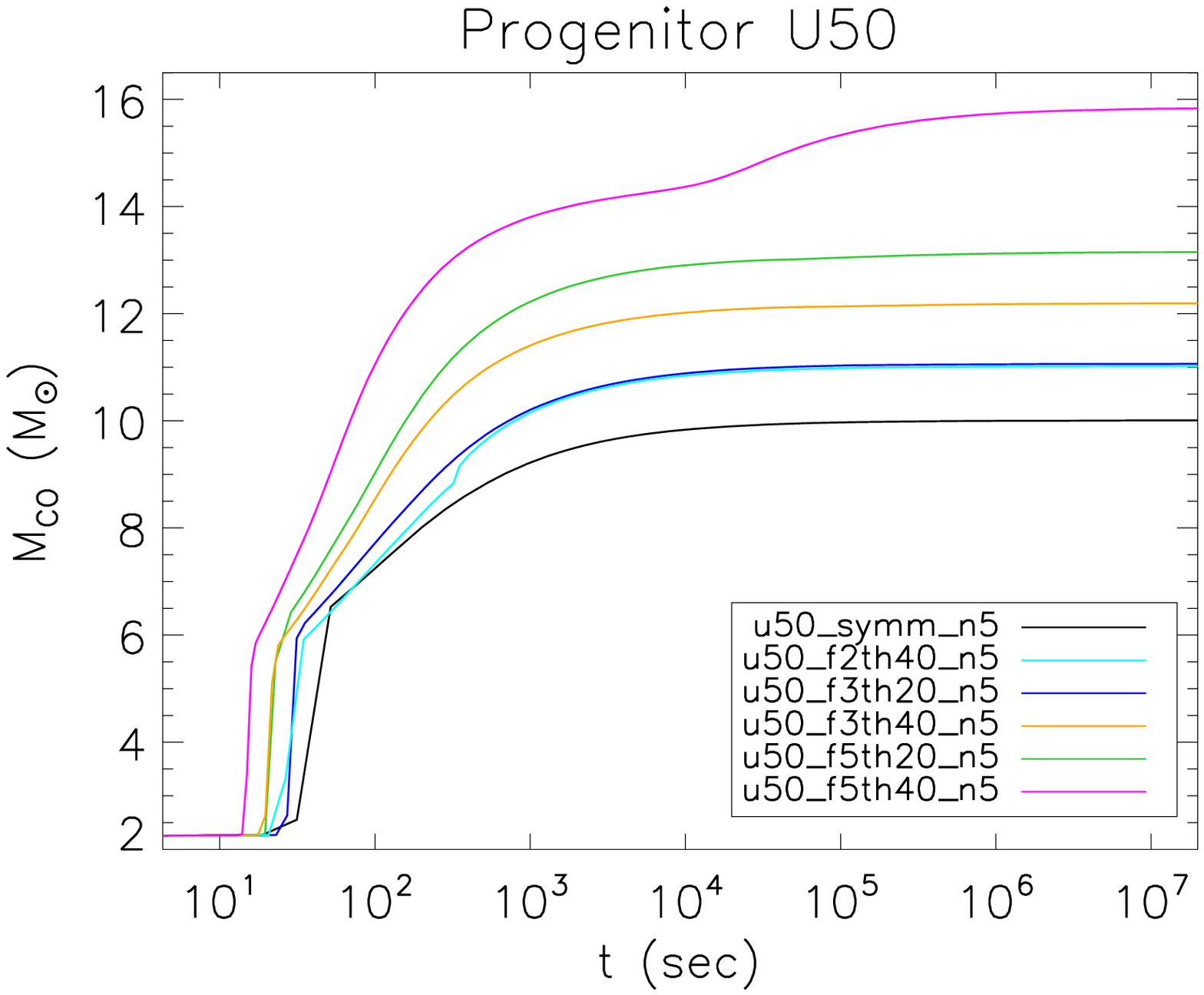}
\plottwo{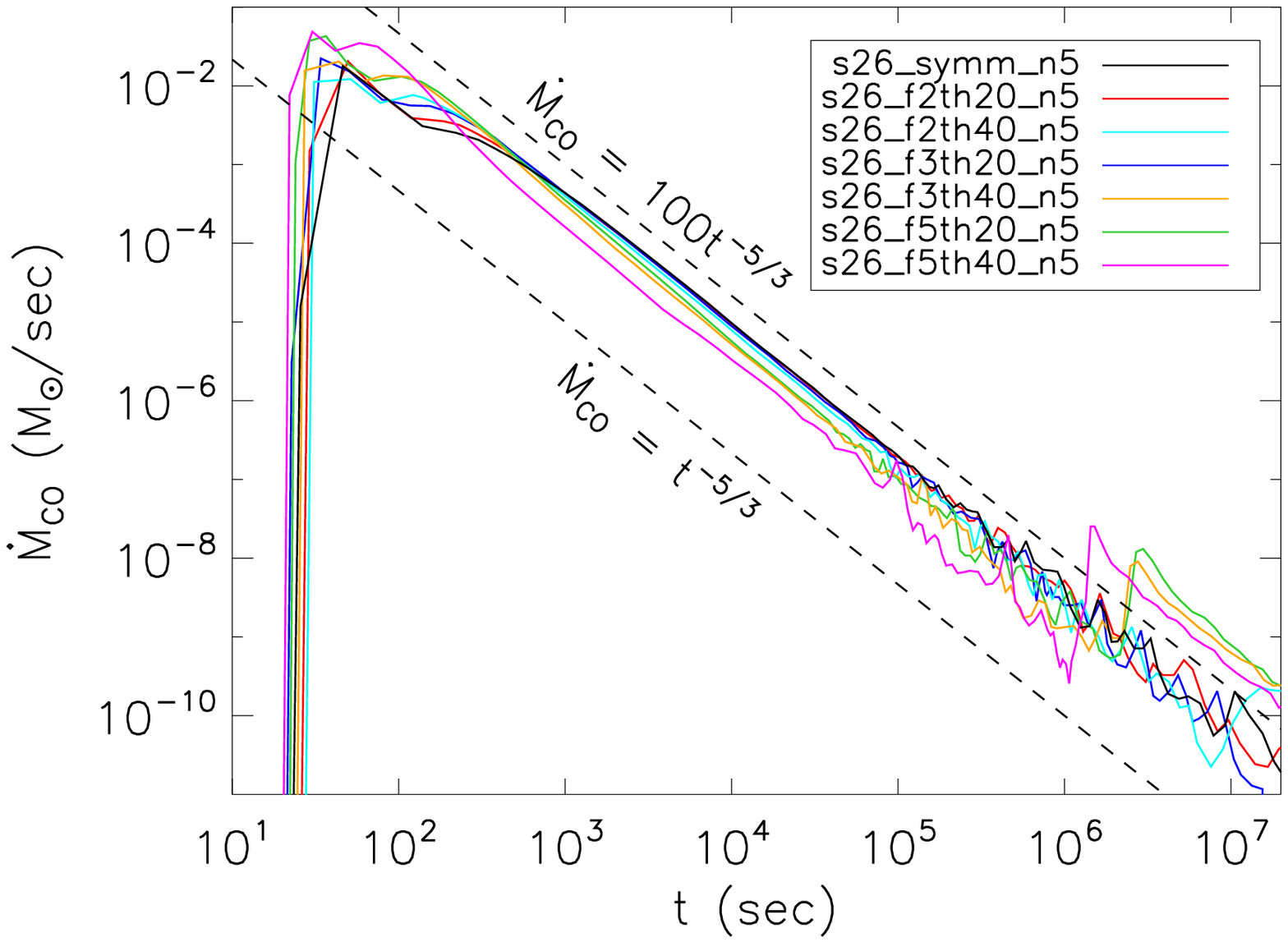}{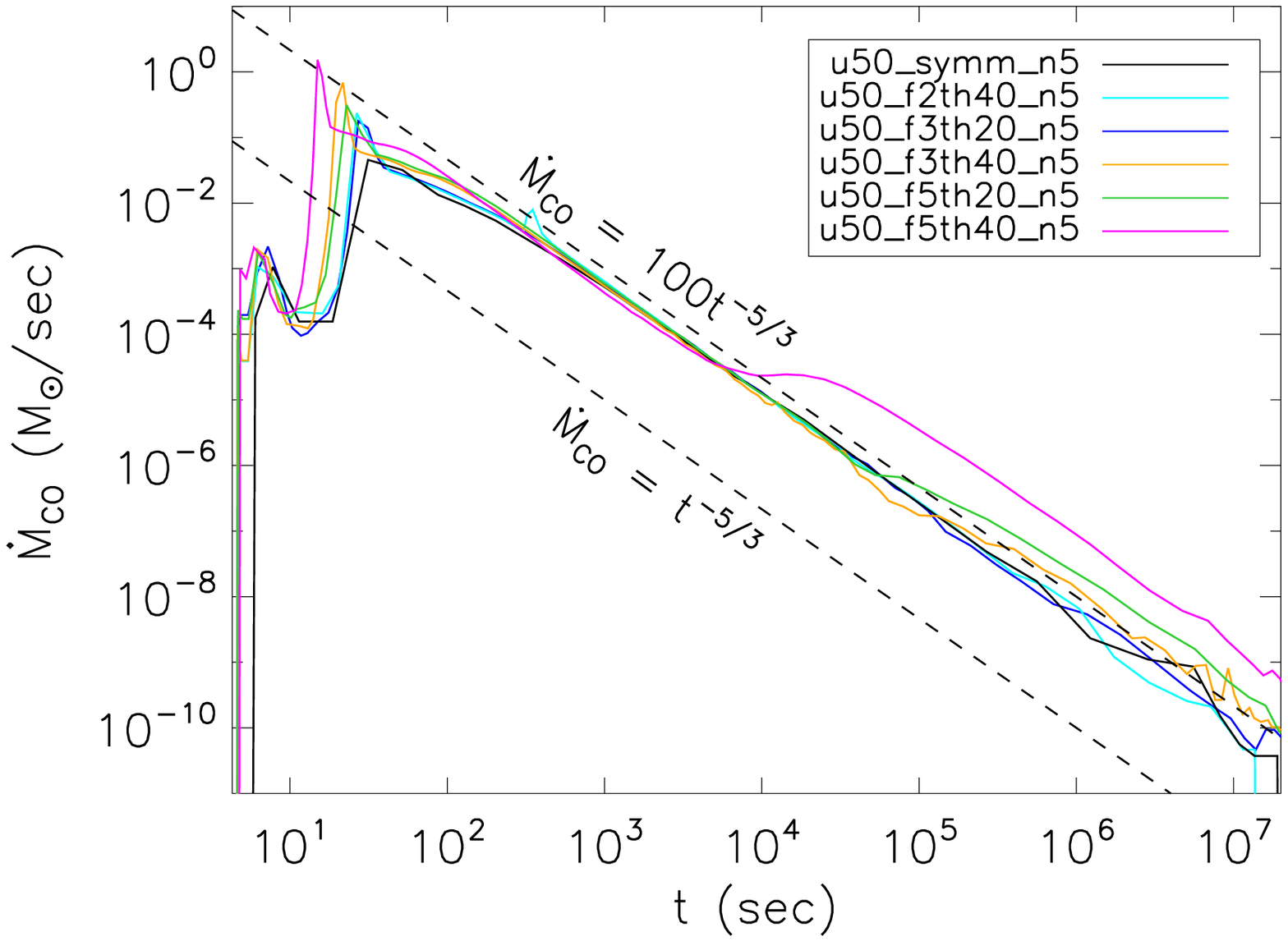}
\caption{
The mass ($M_{co}$, top panels) and accretion rate ($\dot{M}_{co}$, bottom panels) of the central compact object vs. time (t)
elapsed since the start of the explosion for simulations with explosion energies being artificially weakened ($\eta = 0.5$).
After $t \approx 100$\,s, the rates of fallback ($\equiv \dot{M}_{co}$) in most of these simulations decay as $t^{-5/3}$,
which means fallback occurs solely due to prompt fallback mechanism.
In simulations s26\_f3th40\_n5, s26\_f5th20\_n5 and s26\_f5th40\_n5, u50\_f5th20\_n5 and u50\_f5th40\_n5,
the fallback rates at late times either increase by one order of magnitude or stay roughly constant for a short period of time.
This is because the reverse shocks in those simulations produce extra fallback.}
\label{fig:mco_plot2}
\end{center}
\end{figure}

\begin{figure}
\epsscale{1.0}
\begin{center}
\plottwo{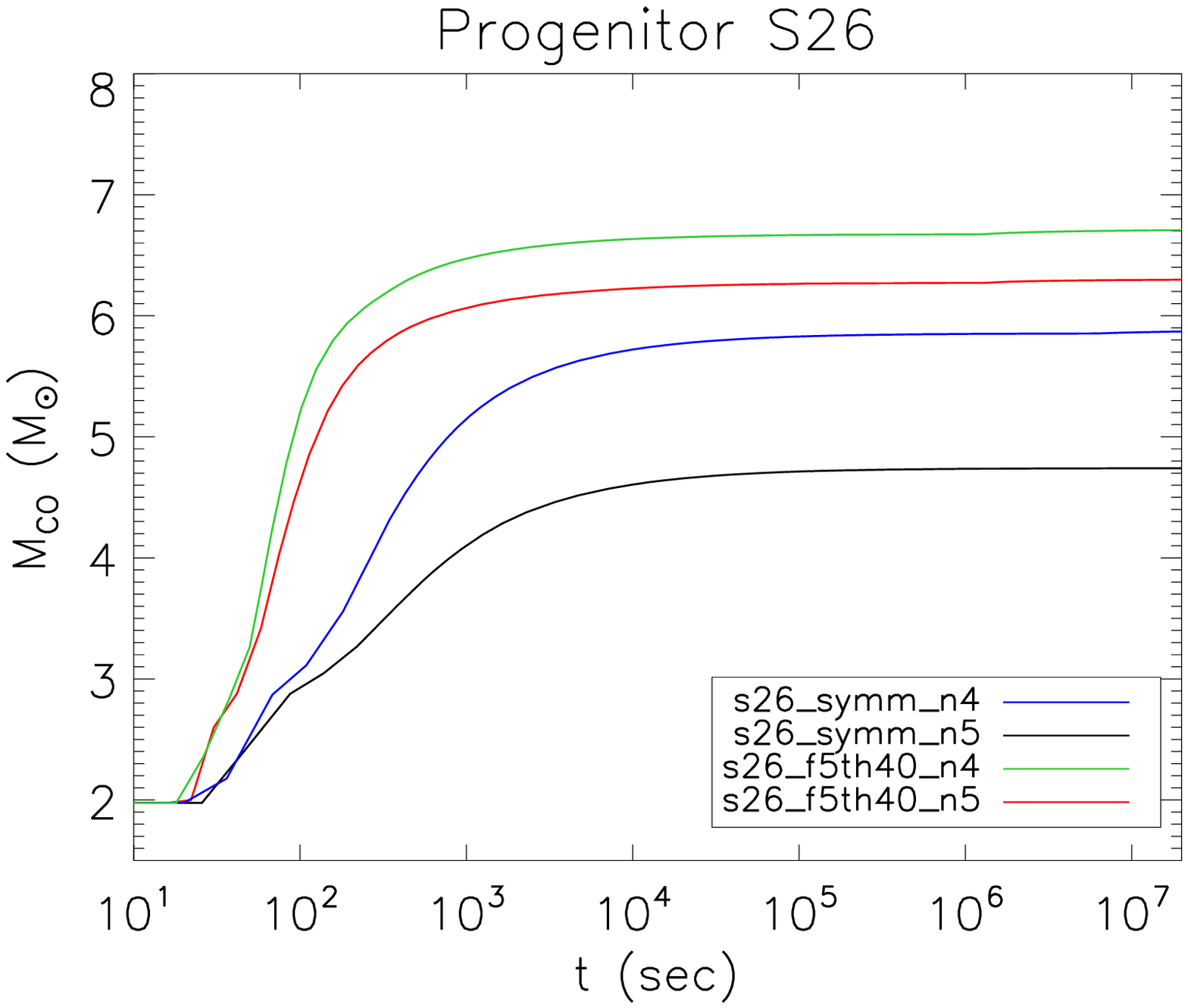}{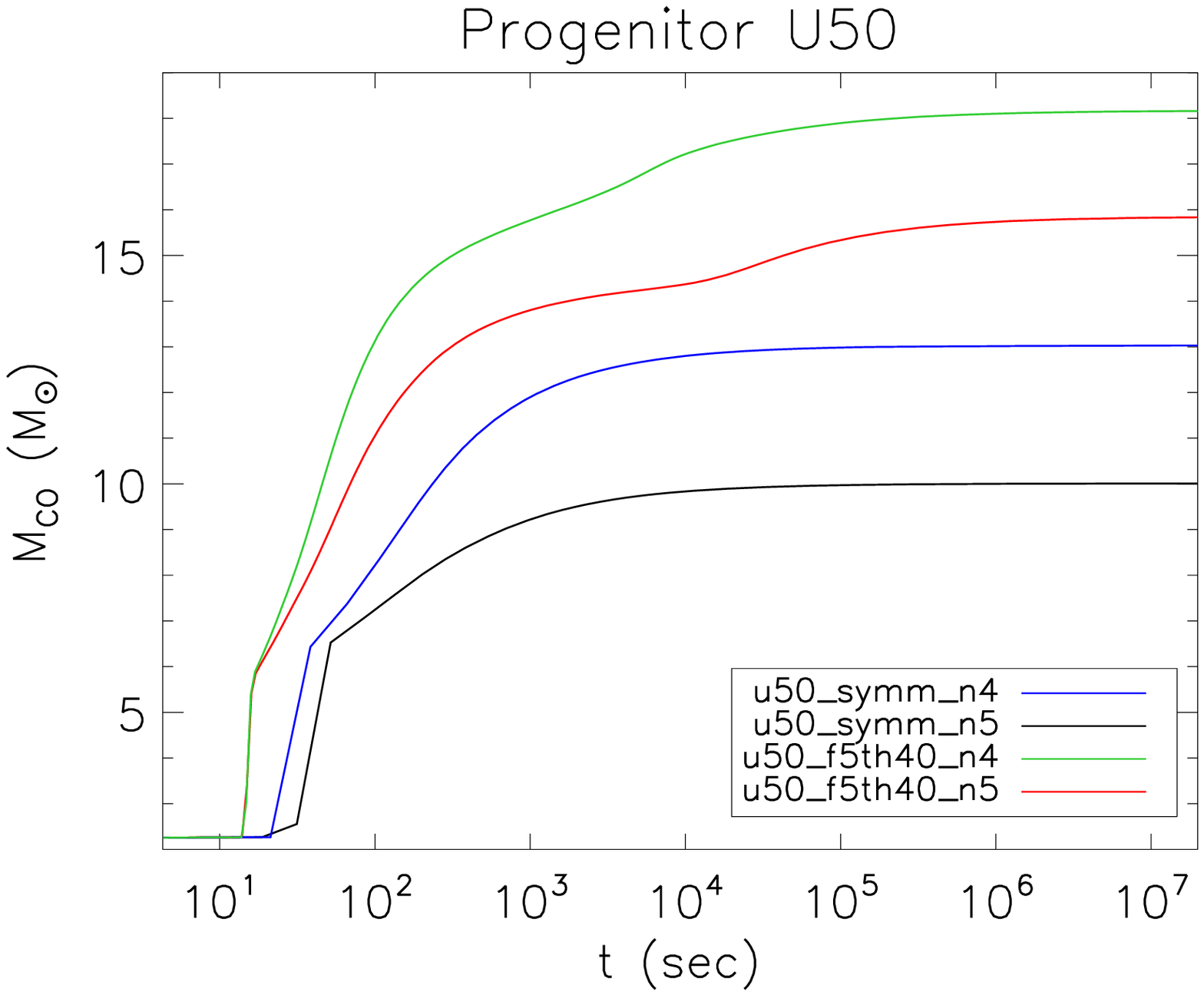}
\plottwo{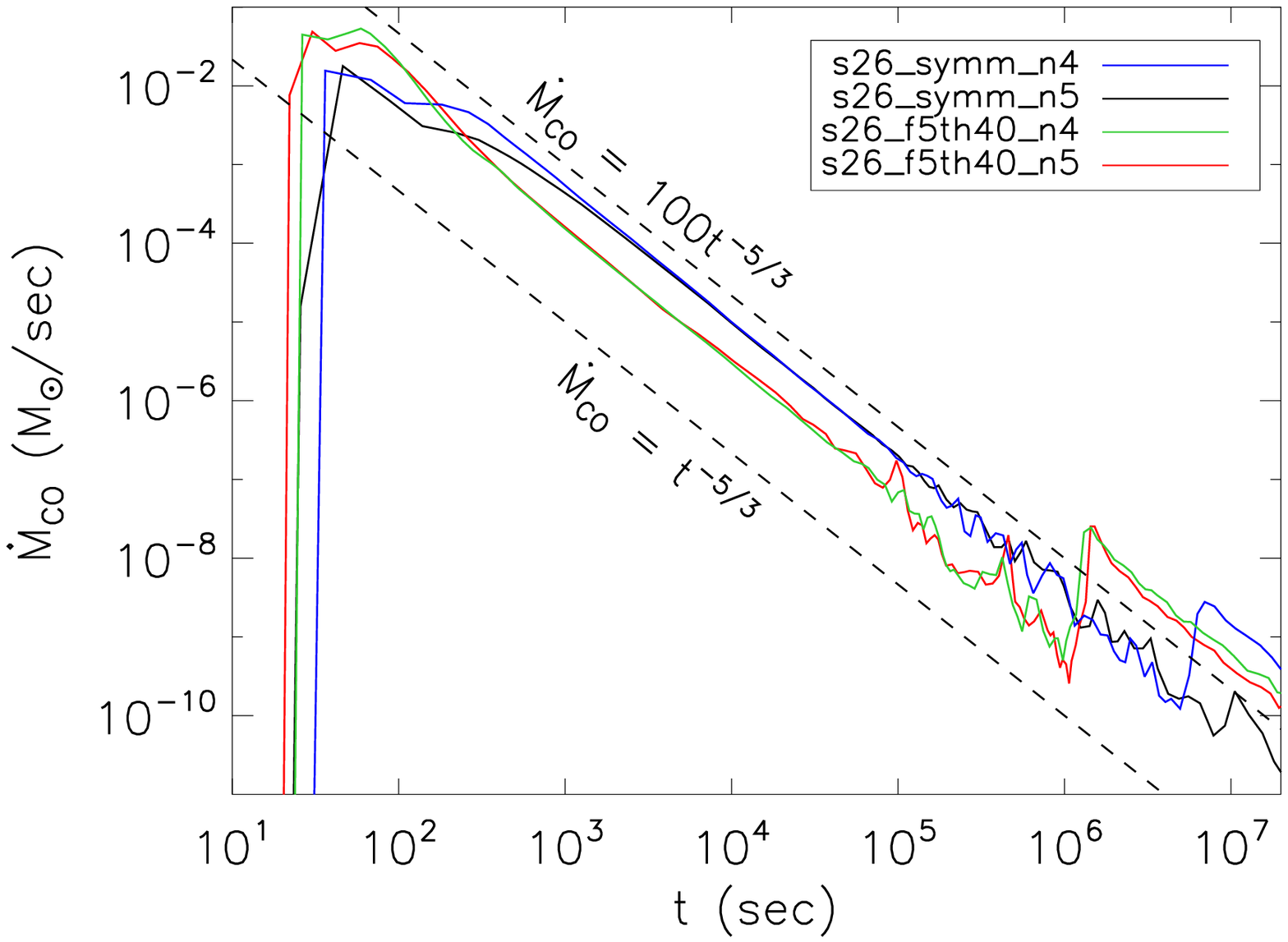}{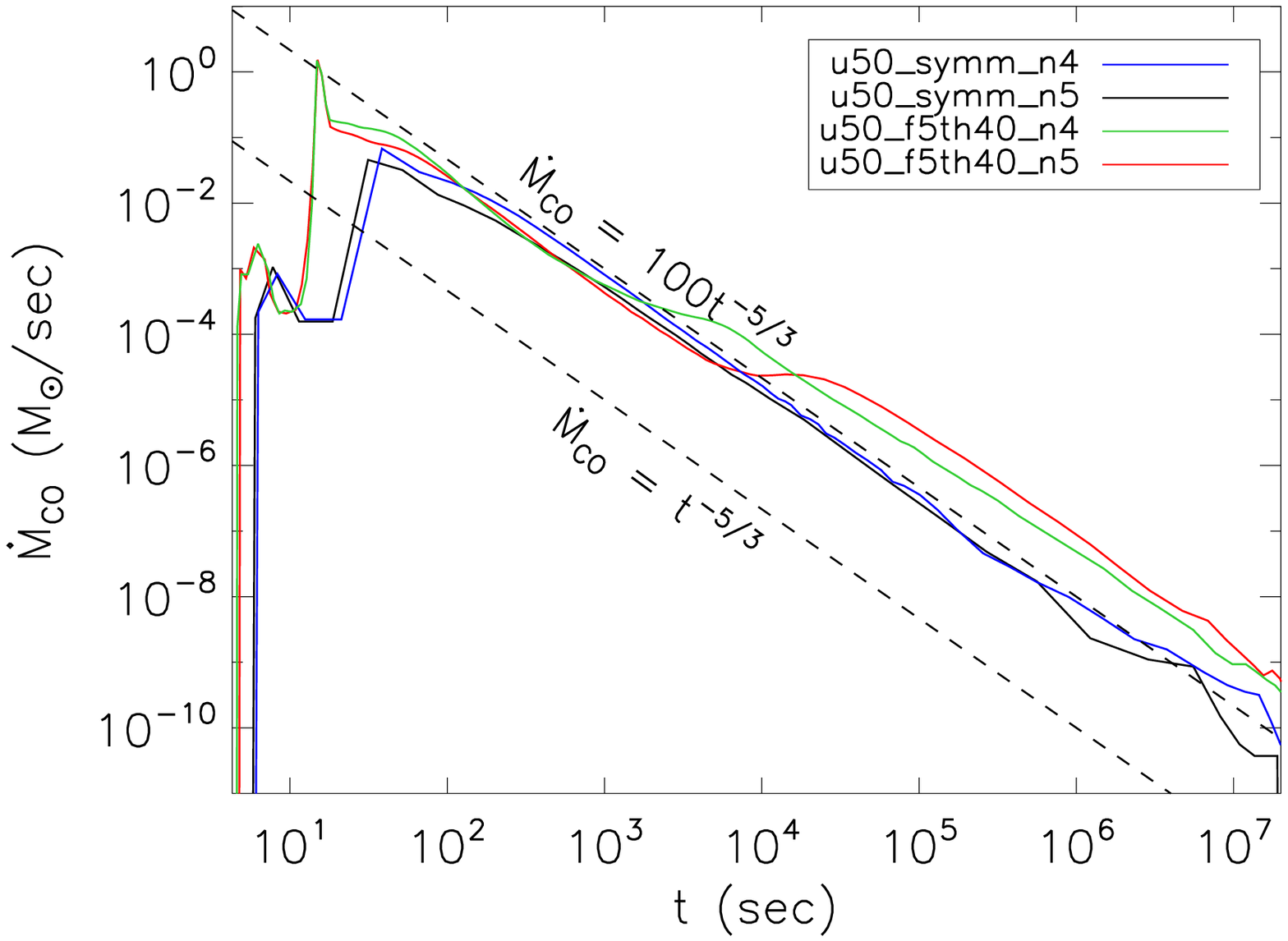}
\caption{
The mass ($M_{co}$, top panels) and accretion rate ($\dot{M}_{co}$, bottom panels) of the central compact object vs. time (t)
elapsed since the start of the explosion for simulations of symmetric and most asymmetric ($f = 5$, $\Theta = 40$ deg) explosions,
with artificially weakened explosion energy ($\eta = 0.4$, 0.5).
In all of these simulations, the rates of fallback ($\equiv \dot{M}_{co}$) decay as $t^{-5/3}$ after $t \approx 100$\,s,
as fallback is primarily produced by the prompt fallback mechanism.
At late times when the reverse shock produces extra fallback (except in simulations s26\_symm\_n5 and u50\_symm\_n5),
the fallback rates either increase by one order of magnitude or stay roughly constant for a short period of time.}
\label{fig:mco_plot3}
\end{center}
\end{figure}

By artificially decreasing the total kinetic and internal energy of the post-shock material by half ($\eta = 0.5$),
we find that the reverse shock in a few runs of asymmetric explosions causes additional fallback.
For the progenitor S26, the reverse shock in simulations of highly asymmetric explosions (s26\_f3th40\_n5, s26\_f5th20\_n5 and s26\_f5th40\_n5)
reaches the center (i.e. the inner absorbing boundary) at $t \approx 12$\,d.
As illustrated in Figure~\ref{fig:mco_plot2}, when the reverse shock arrives at the center, the fallback rates increase by a factor of $\sim10$.
Despite these dramatic increases in fallback rates, the amount of extra fallback due to the reverse shock is $< 0.05\,M_\sun$.
This is because those enhanced fallback rates are very low compared to the peak of fallback, and they decay rapidly.

With the same artificial decrease in the energy of the post-shock material,
the reverse shock also affects the amount of fallback in two asymmetric explosions of the progenitor U50
(u50\_f5th20\_n5 and u50\_f5th40\_n5).
In $t \approx 1$--10\,hr, the reverse shock in these simulations propagates close to the inner absorbing boundary.
Then, material from the reserve shock front starts to fall across the inner boundary, causing additional fallback and gradually dissolving the
reverse shock into the flow of fallback material.
At the peak of this additional fallback due to the reverse shock, the accretion rate of the central compact object stays roughly constant for 3--6\,hr
(see Figure~\ref{fig:mco_plot2}).
In the simulation u50\_f5th20\_n5, the reverse shock only leads to a small increase in the amount of fallback material.
However, in the simulation u50\_f5th40\_n5, the reverse shock produces an extra 1\,$M_\sun$ fallback.

To further study the dependence of the fallback nature on the explosion energy, we perform simulations of symmetric and the most
asymmetric ($f = 5$ and $\Theta = 40$~deg) explosions of our progenitor stars with an energy scaling factor $\eta = 0.4$.
Unlike our previous simulations,
the reverse shocks in all of these runs (s26\_symm\_n4, s26\_f5th40\_n4, u50\_symm\_n4 and u50\_f5th40\_n4) reach the center,
producing different amount of additional fallback.
When comparing the simulations s26\_f5th40\_n4 and u50\_f5th40\_n4 to s26\_f5th40\_n5 and u50\_f5th40\_n5, respectively,
we find that the reverse shock in a less energetic explosion of the same progenitor arrives at the center earlier (see Figure~\ref{fig:mco_plot3}).
However, the amount of fallback due to the reverse shock in these less energetic explosions does not seem to be significantly increased.
The increases in the amount of fallback material in those runs are mainly caused by the decreases in explosion energies.

\subsection{Convergence Test}
\label{subsec:convergence_test}

\begin{figure}
\epsscale{0.6}
\begin{center}
\plotone{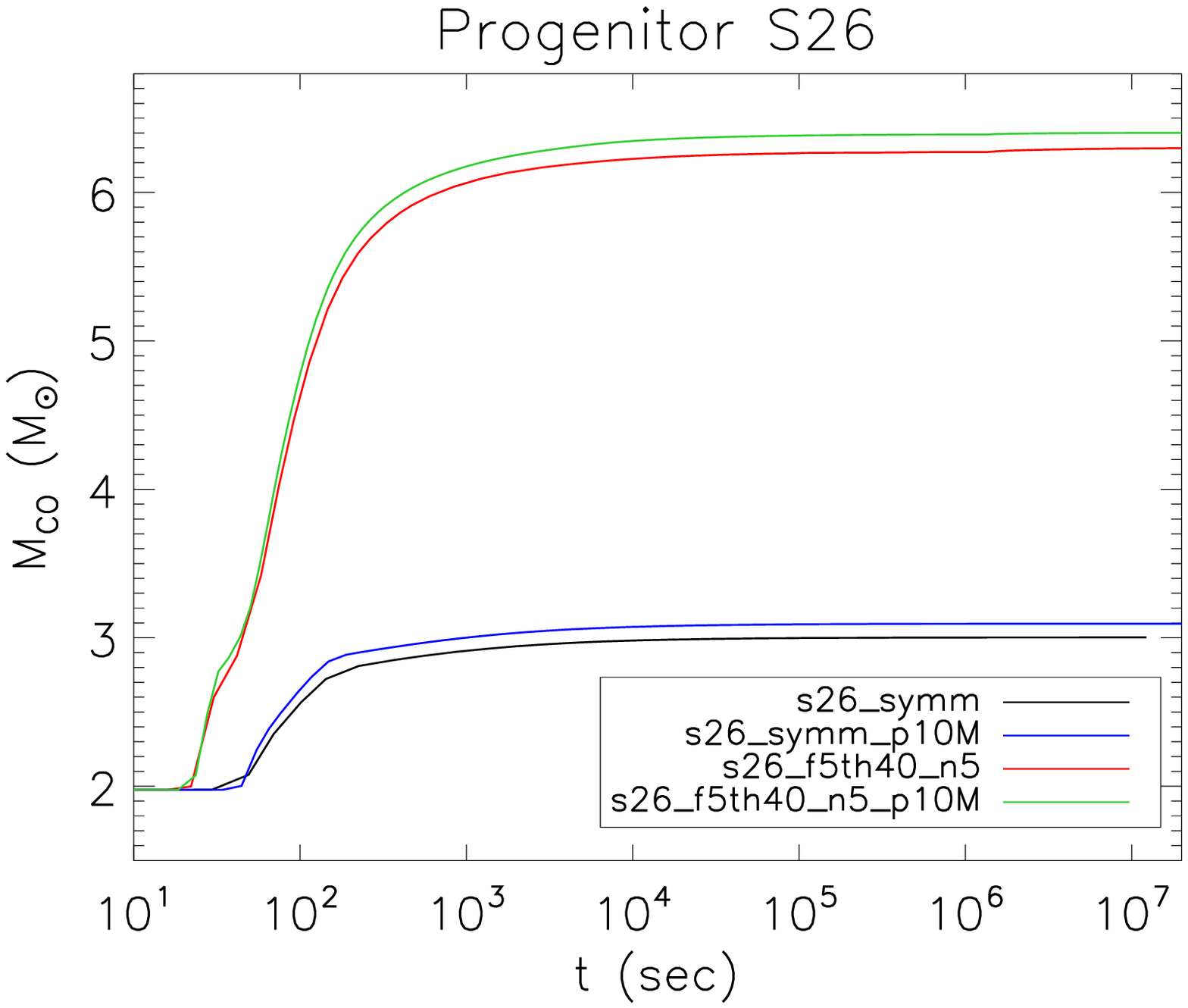}
\plotone{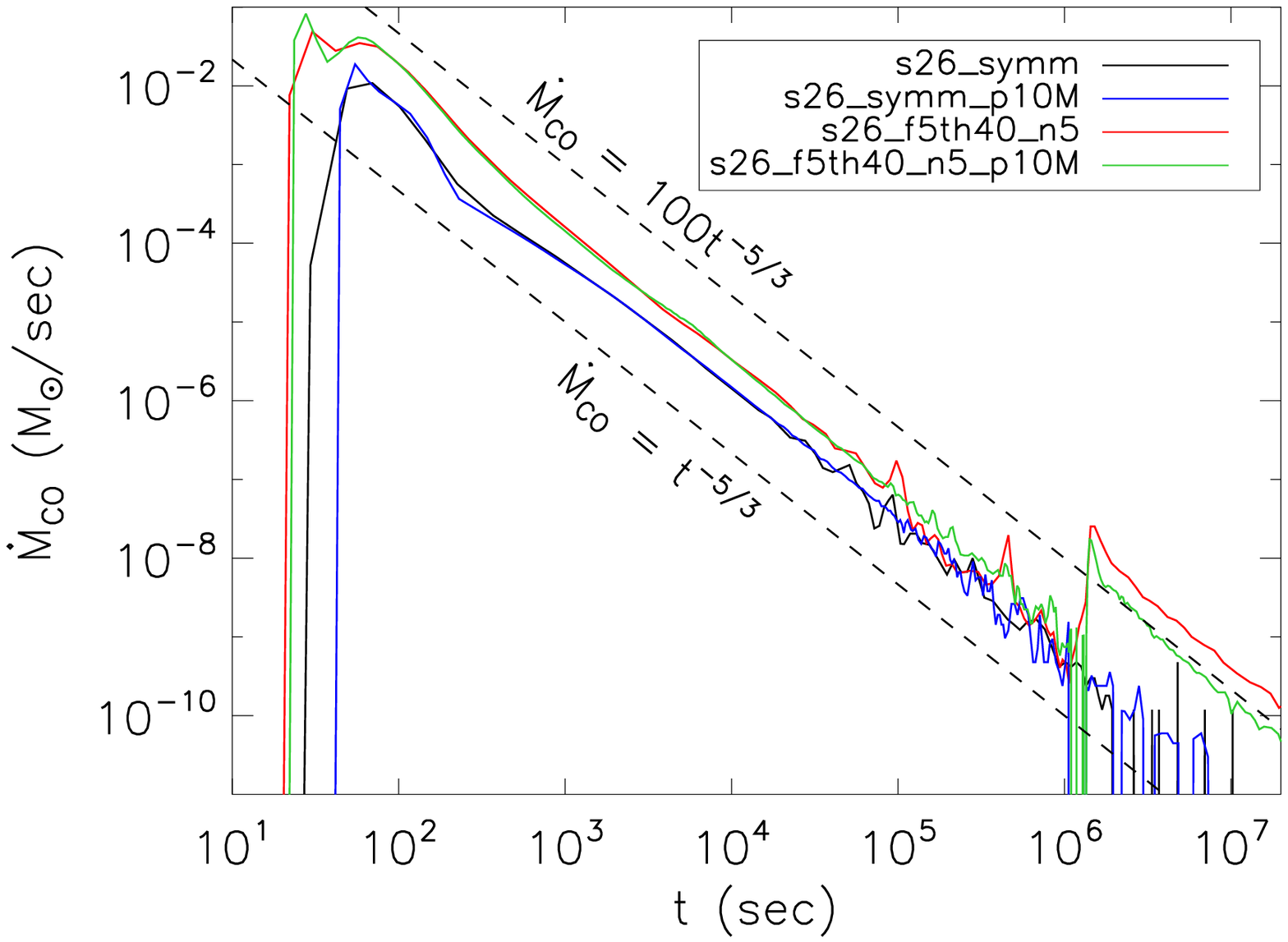}
\caption{The mass ($M_{co}$, top panels) and accretion rate ($\dot{M}_{co}$, bottom panels) of the central compact object vs. time (t)
elapsed since the start of the explosion for simulations of different resolution.
Apart from the number of SPH particles used,
the simulation s26\_symm and s26\_f5th40\_n5 have the same set up as the simulation s26\_symm\_p10M and s26\_f5th40\_n5\_p10M, respectively.
We find that the nature and evolution of fallback in these simulations does not change with the resolution.}
\label{fig:convergence_test}
\end{center}
\end{figure}

To see whether our results depend on the resolution of our 3D simulations, we run two explosion simulations
(s26\_symm\_p10M and s26\_f5th40\_n5\_p10M) of the progenitor S26 using 9.5 million SPH particles.
Here, the SPH particle masses in the initial 3D configuration range from 3.4 to $29 \times 10^{-7} M_\odot$, with an average of $11 \times 10^{-7} M_\odot$.
Apart from the number of SPH particles used, these two simulations have the same set up as the corresponding simulations of lower resolution (s26\_symm
and  s26\_f5th40\_n5, respectively).
The final masses of the central compact objects in both simulations of higher resolution are slightly more massive (see Figure~\ref{fig:convergence_test}).
This is because the density spike due to our 1D energy deposition model, which produces the fallback in the first 100\,s after the launch of the explosion,
is better resolved.
After $t = 100$\,s, the fallback rates of the 9.5 million particles simulations agree well with those of the corresponding 1.2 million particles
simulations.
In both simulation of the symmetric explosion, the reverse shock freezes in the expanding ejecta and never gets close to the center.
At $t \approx 0.5$\,yr when these simulations are terminated, the reverse shocks in both simulations have the same radius of $2.2 \times 10^{15}$\,cm,
despite the different number of SPH particles located at radii less than the reverse shock:
$1.6 \times 10^4$ and $1.3 \times 10^5$ for the simulation of lower and higher resolution, respectively.
In the simulations of the asymmetric explosion with weakened explosion energy ($\eta = 0.5$), the reverse shock in both 1.2 and 9.5 million particles
simulation reaches the center at the same time (see Figure~\ref{fig:convergence_test}).
Both simulations of different resolution show that the amount of fallback resulted from the reverse shock is negligible.
As a result, our choice of running 3D simulations with 1.2 million particles is good enough for this study.

\section{Discussion}

\label{sec:discussion}

We have studied the fallback in 3D for two massive progenitor stars with a range of explosion asymmetries and energies.
Our progenitor models are a 26\,$M_\odot$ star of solar metallicity (S26) and a 50\,$M_\odot$ star of zero metallicity (U50),
which were created by \cite{woosley...02}.
In all of our runs, fallback peaks in the first 10--100\,s after the launch of the explosion.
After this peak, the fallback accretion rate decays with time at a rate proportional to $t^{-5/3}$
(see Figure~\ref{fig:mco_plot1}, \ref{fig:mco_plot2}--\ref{fig:convergence_test}).
All of our simulations show that fallback occurs primarily due to the prompt fallback mechanism \citep{colgate71, fryer99}.
Indeed, in most cases, the reverse shock freezes in the flow of the expanding ejecta and does not drive fallback at all.

\begin{figure}
\begin{center}
\plotone{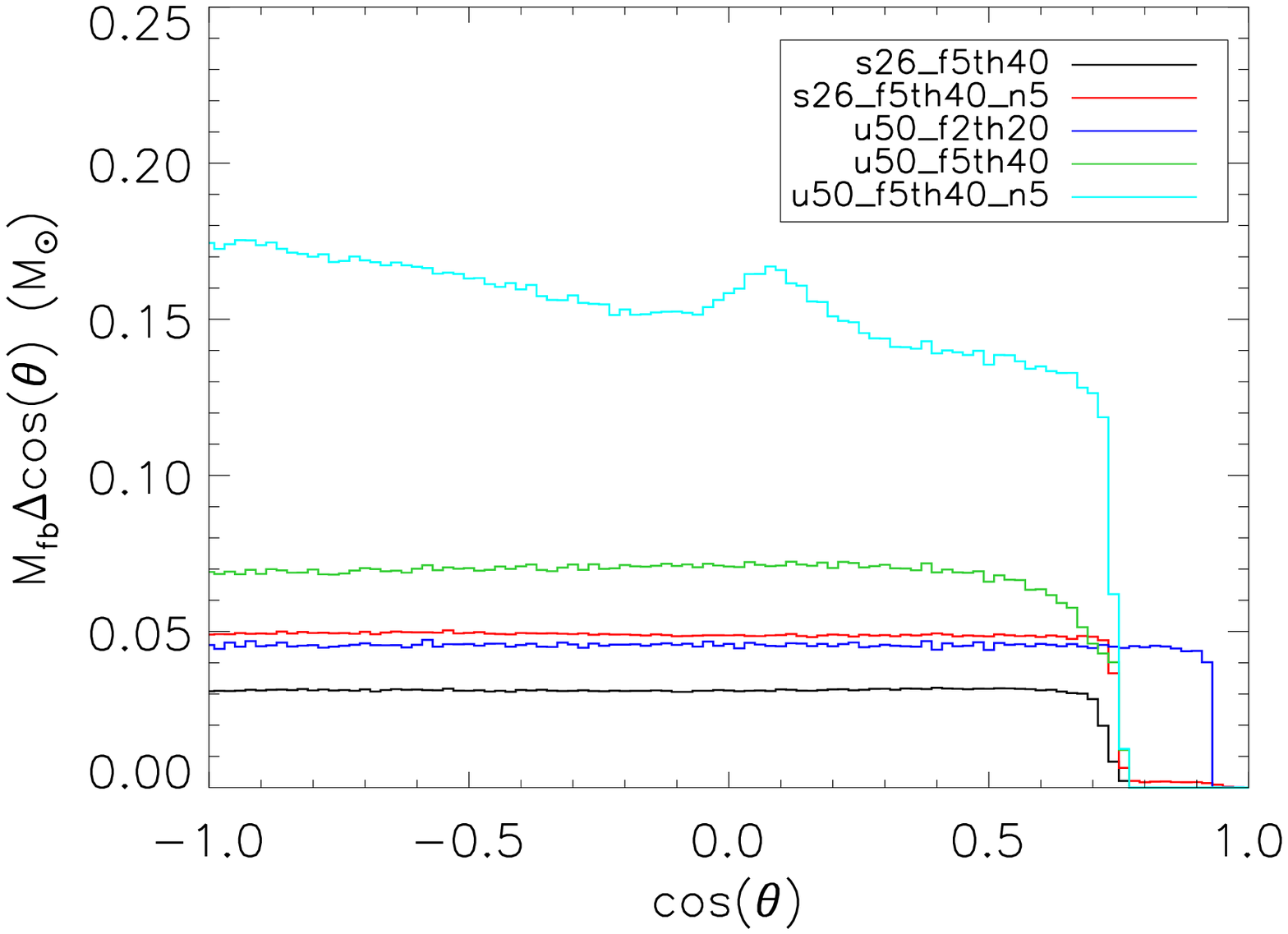}
\caption{The distributions of the fallback origin ($M_{fb} \Delta \cos \theta)$ as functions of $\cos(\theta)$ for several runs of asymmetric explosions.
Here, $\theta$ is the polar coordinate.
The axes of the enhanced cones of explosions in these runs all point along the direction of $\theta = 0$.
These distribution functions have the shape of step functions because our injected asymmetries are step functions of $\cos \theta$.
Nevertheless, the bulk of fallback occurs outside the cone of enhanced explosion, where the explosion is weakest.}
\label{fig:fallback_dist}
\end{center}
\end{figure}

In our simulations of asymmetric explosions, we note that the bulk of the fallback occurs where the explosion is weakest,
i.e. outside the cone of enhanced explosion (see Figure~\ref{fig:fallback_dist}).
Based on the current understanding of core-collapse supernova, the explosion is weakest where the angular momentum is highest
\citep[e.g.][]{fryerheger00}.
Although we did not include angular momentum in our study, in most cases, the fallback material will have a higher average specific angular
momentum than the progenitor star.

Our study shows that the amount of fallback depends both on the explosion energy and its asymmetry.
For our strongest explosions ($\eta = 1$), the level of asymmetry leads to a range of fallback from 1.0--2.7\,$M_\odot$ for our progenitor S26
and 4.3--6.0\,$M_\odot$ for our progenitor U50, with larger asymmetries producing more fallback (see Table\;\ref{tab:simulations_summary}).
The resulting remnant masses for our progenitor S26 in the simulations with $\eta = 1$ lie exactly in the reported black hole mass gap (or paucity range)
of 3--5\,$M_\odot$ \citep{bailyn...98, ozel...10, farr...11}.
An observed mass gap could be explained in a number of ways \citep[see e.g.][]{fryer99, fryer12, belczynski12, kreidberg...12, kochanek13}.
The simplest one is that the explosion energies are weaker for these stars (e.g. see Figure\,\ref{fig:mco_plot2}).
However, we note that such a mass gap also allows strong explosions as long as they are sufficiently asymmetric (see Figure\;\ref{fig:mco_plot1}).

The explosions in this study all produce high fallback masses (>\,1\,$M_\odot$), with peak fallback rates in the range of 0.01--1\,$M_\odot/s$.
Especially in the runs of the progenitor U50, we note that the peak of the fallback in each 3D simulation appears as a spike in the early-time (10--100\;s) fallback rate.
This is due to the fact that our explosions are driven by the 1D energy-deposition model, which allows material to pile up above the energy injection region,
producing double-shocked material (see \S\;\ref{sec:1Dmodel}).
In multi-dimensional explosion models, it is likely that part of this pile up material will flow through the energy injection region,
producing a smoother accretion profile at early times.
Even so, the fallback will still produce some spikes in the compact object accretion rate.
Hence, our 1D energy-deposition model gives an upper limit on the strength of this spike.
Nevertheless, the early-time fallback rates in this study are in general high enough to cause a neutrino flux on par with that of the cooling neutron star \citep{fryer09}.
We will have to disentangle these two neutrino sources to study neutron star equations of state.
Especially at late times of an explosion, the neutrino emission can be used to estimate the accretion rate, providing a window into the nature
of the collapse of a neutron star to a black hole.
As the fallback material is likely to form a disk around the compact object, detailed calculations will be needed to better follow this collapse
process.

Fallback can also play a role in the r-process nucleosynthetic yields in systems with either a neutron star or black hole compact remnant.
In systems with a neutron star compact remnant, the fallback material releases sufficient energy to drive outflows.
The yields from these outflows could dominate the r-process yields in the early universe \citep{fryeretal06}.
Alternatively, fallback can also form r-process yields when the compact remnant is a black hole, as long as the fallback has enough angular
momentum to form a disk around the black hole.
The wind from this accretion disk can eject additional $^{56}$Ni and, in some cases, r-process elements \citep{surman11}.

Models with considerable fallback can explain some observations of peculiar supernovae \citep[e.g.][]{foley...09, quimby...11, galyam12}.
\cite{fryer...09} and \cite{moriya...10} argued that these models could produce dim supernova explosions.
However, the outbursts seen in \cite{fryeretal06}, which are driven by fallback accretion, can also provide additional energy to the supernova explosion.
\cite{dexter13} argued that such outbursts could explain super-luminous supernovae.
As such outbursts are driven by energy released from the accretion onto a compact remnant, we term this feedback mechanism as ``accretion heating''
mechanism.
In the limit that the specific angular momentum of the fallback material is low, \cite{fryeretal06} studied the fallback onto a proto-neutron star and
argued that $\sim\,25\%$ of the fallback material will be ejected as outflows at velocities of $\sim\,10^4$\,km/s.
Using our peak fallback rates and assuming that the outflow material deposits all of its kinetic energy to the ejecta, we find that the explosion luminosities
produced by this ``accretion heating'' mechanism can be as high as $10^{48}$--$10^{50}$\,erg/s.

\begin{figure}
\begin{center}
\plotone{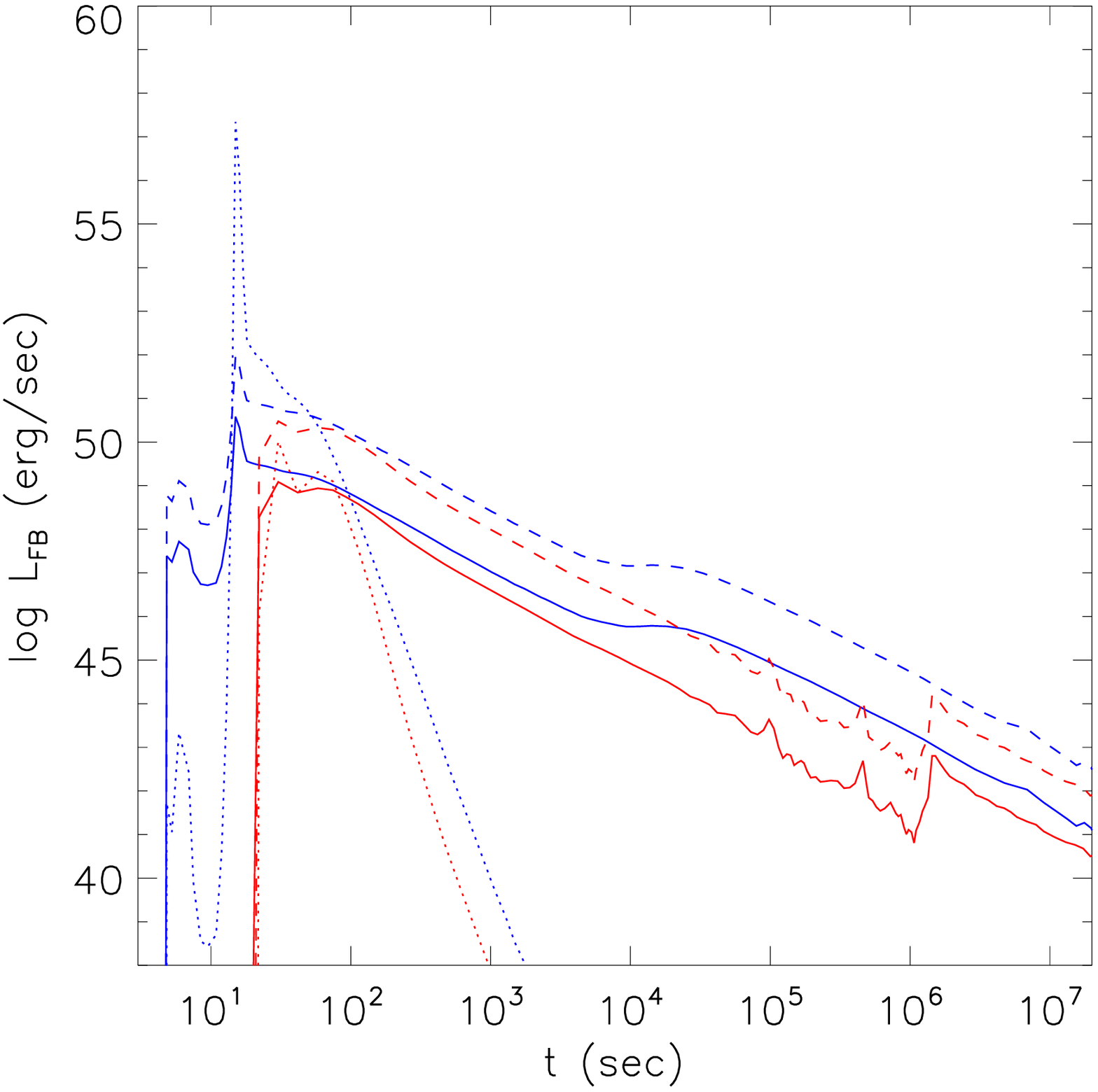}
\caption{
The time evolution of our estimates on the explosion luminosity due to fallback for two selected simulations: s26\_f5th40\_n5 (red) and u50\_f5th40\_n5 (blue).
We consider fallback explosion luminosity powered by three mechanisms: ``accretion heating'' (solid line), neutrino annihilation (dotted line),
and Blandford-Znajek emission (dashed line).
Our estimates show that fallback can potentially lead to large amount of energy deposition to the ejecta, powering super-luminous supernovae. 
}
\label{fig:fallback_Lexp}
\end{center}
\end{figure}

If the specific angular momentum of the fallback material is large enough to form an accretion disk, we follow \cite{zhang&fryer01} and estimate
the explosion luminosity assuming both neutrino annihilation and Blandford-Znajek mechanisms.
For the neutrino annihilation mechanism, we use the formula from \cite{zhang&fryer01}:
\begin{equation}
\log \left [L_{\nu,\bar\nu} (erg/s) \right ] \approx 43.6 + 4.89 \log \left ( \frac{\dot{M}_{co}}{0.01 M_\odot/s} \right ) + 3.4a_{co}.
\end{equation}
We set the spin parameter ($a_{co}$) of the compact object to 0.9 for our estimates.
We find that $L_{\nu,\bar\nu}$ at the peak of fallback in our simulations is in the range of $10^{47}$--$10^{57}$\,erg/s.
For Blandford-Znajek emission, a comparable equation was given by \cite{popham99}:
\begin{equation}
\log \left [L_{BZ} (erg/s) \right ] \approx 50.0 + 2\log \left [ a_{co} \biggl (\frac{M_{co}}{3 M_\odot} \biggr ) \biggl (\frac{B_{mag}}{10^{15} G} \biggr )\right ],
\end{equation}
where $B_{mag}$ is the strength of the magnetic field.
Assuming that the magnetic energy density is equal to the kinetic energy density \citep{zhang&fryer01, fryer...13}, we can write
\begin{equation}
B^2_{mag} = 4\pi\rho v^2 \approx \dot{M}_{co} c/r_g^2,
\end{equation}
where $c$ is the speed of light and $r_g \equiv 2GM_{co}/c^2$ is the radius of the event horizon.
Again, we set $a_{co}$ to 0.9 for our estimates.
At the peak of fallback in our simulations, $L_{BZ}$ falls within the range of $10^{50}$--$10^{52}$\,erg/s.
As a result, our estimated upper limits of fallback explosion luminosity show that fallback can potentially deposit a large amount of energy to the ejecta,
producing super-luminous supernovae.
In Figure\;\ref{fig:fallback_Lexp}, we illustrate the time evolution of the fallback explosion luminosity given by the above mentioned mechanisms
for two selected simulations.
Those simulations have fallback driven by both prompt fallback and reverse shock mechanisms.
It is clear that the fallback explosion luminosity peaks at the same time as the fallback rate.
Also, the fallback driven by reverse shock deceleration plays a minor role in powering fallback explosion luminosity.

Finally, fallback can also connect to gamma-ray bursts.
The newly formed neutron star in most of our simulations will accrete enough fallback material and further collapse to a black hole.
The higher specific angular momentum in this fallback material makes it easier to form a clean accretion disk, which can potentially lead to a
gamma-ray burst.
These so-called ``type II'' collapsars \citep{macfadyen01,heger03} have slightly different features than direct-collapse collapsars
\citep{woosley93, macfadyen=woosley99}.
The accretion rate of a ``type II'' collapsar is typically an order of magnitude lower than that of a direct-collapse collapsar,
meaning that neutrino-annihilation mechanisms are unlikely to operate \citep{popham99}.
The weak explosion followed by fallback and a second outburst will eject far less $^{56}$Ni than normal hypernova/gamma-ray bursts.
Hence, a ``type II'' collapsar produces a unique supernova associated with a gamma-ray burst \citep{fryer06,fryer07}.
However, how much the yields vary will require more detailed multi-dimensional calculations which we defer to a later paper.

\acknowledgements
This project was done under the auspices of the National Nuclear Security Administration of the U.S. Department of Energy, and supported by its contract
DE-AC52-06NA25396 at Los Alamos National Laboratory.
It was also partially supported by a Simons Fellowship in Theoretical Physics awarded to VK.
TW acknowledges support from NSF Grant AST-0908930 and a CIC/Smithsonian pre-doctoral fellowship, as well as the hospitality of
Harvard-Smithsonian Center for Astrophysics.


\begin{thebibliography}{}

\bibitem[Bailyn et al.(1998)]{bailyn...98} Bailyn, C.~D., Jain, R.~K., Coppi, P., \& Orosz, J.~A.\ 1998, \apj, 499, 367 
\bibitem[Belczynski et al.(2012)]{belczynski12} Belczynski, K., Wiktorowicz, G., Fryer, C.~L., Holz, D.~E., \& Kalogera, V.\ 2012, \apj, 757, 91 
\bibitem[Benz(1984)]{benz84} Benz, W.\ 1984, \aap, 139, 378
\bibitem[Benz(1988)]{benz88} Benz, W.\ 1988, Computer Physics Communications, 48, 97
\bibitem[Benz et al.(1989)]{benz...89} Benz, W., Thielemann, F.-K., \& Hills, J.~G.\ 1989, \apj, 342, 986
\bibitem[Benz(1990)]{benz90} Benz, W.\ 1990, , in The Numerical Modeling of Stellar Pulsation, ed. J. R. Buchler (Dordrecht: Kluwer), 269
\bibitem[Blondin et al.(2003)]{blondin...03} Blondin, J.~M., Mezzacappa, A., \& DeMarino, C.\ 2003, \apj, 584, 971 
\bibitem[Bruenn et al.(2013)]{bruenn...13} Bruenn, S.~W., Mezzacappa, A., Hix, W.~R., et al.\ 2013, \apjl, 767, L6 

\bibitem[Chevalier(1989)]{chevalier89} Chevalier, R.~A.\ 1989, \apj, 346, 847 
\bibitem[Chiosi \& Summa(1970)]{chiosi=summa70} Chiosi, C., \& Summa, C.\ 1970, \apss, 8, 478
\bibitem[Chiosi et al.(1978)]{chiosi...78} Chiosi, C., Nasi, E., \& Sreenivasan, S.~R.\ 1978, \aap, 63, 103
\bibitem[Chiosi \& Maeder(1986)]{chiosi=maeder86} Chiosi, C., \& Maeder, A.\ 1986, \araa, 24, 329
\bibitem[Colgate(1971)]{colgate71} Colgate, S.~A.\ 1971, \apj,  163, 221 

\bibitem[De Beck et al.(2010)]{debeck...10} De Beck, E., Decin, L., de Koter, A., et al.\ 2010, \aap, 523, A18
\bibitem[Demorest et al.(2010)]{demorest...10} Demorest, P.~B., Pennucci, T., Ransom, S.~M., Roberts, M.~S.~E., \& Hessels, J.~W.~T.\ 2010, \nat, 467, 1081
\bibitem[Dexter \& Kasen(2013)]{dexter13} Dexter, J., \& Kasen, D.\ 2013, \apj, 772, 30 
\bibitem[Diehl et al.(2012)]{diehl...12} Diehl, S., Rockefeller, G., Fryer, C.~L., Riethmiller, D., \& Statler, T.~S.\ 2012, arXiv:1211.0525

\bibitem[Ellinger et al.(2012)]{ellinger...12} Ellinger, C.~I., Young, P.~A., Fryer, C.~L., \& Rockefeller, G.\ 2012, \apj, 755, 160
\bibitem[Ellinger et al.(2013)]{ellinger...13} Ellinger, C.~I., Rockefeller, G., Fryer, C.~L., Young, P.~A., \& Park, S.\ 2013, arXiv:1305.4137

\bibitem[Farr et al.(2011)]{farr...11} Farr, W.~M., Sravan, N., Cantrell, A., et al.\ 2011, \apj, 741, 103
\bibitem[Finn(1994)]{finn94} Finn, L.~S.\ 1994, Physical Review Letters, 73, 1878
\bibitem[Foley et al.(2009)]{foley...09} Foley, R.~J., Chornock, R., Filippenko, A.~V., et al.\ 2009, \aj, 138, 376 
\bibitem[Fragos et al.(2009)]{fragos...09} Fragos, T., Willems, B., Kalogera, V., et al.\ 2009, \apj, 697, 1057
\bibitem[Freire et al.(2008)]{freire...08} Freire, P.~C.~C., Ransom, S.~M., B{\'e}gin, S., et al.\ 2008, \apj, 675, 670
\bibitem[Fryer(1999)]{fryer99} Fryer, C.~L.\ 1999, \apj, 522, 413
\bibitem[Fryer \& Heger(2000)]{fryerheger00} Fryer, C.~L., \& Heger, A.\ 2000, \apj, 541, 1033
\bibitem[Fryer \& Kalogera(2001)]{fryerkalogera01} Fryer, C.~L., \& Kalogera, V.\ 2001, \apj, 554, 548 
\bibitem[Fryer et al.(2006a)]{fryer...06} Fryer, C.~L., Rockefeller, G., \& Warren, M.~S.\ 2006a, \apj, 643, 292
\bibitem[Fryer et al.(2006b)]{fryeretal06} Fryer, C.~L., Herwig, F., Hungerford, A., \& Timmes, F.~X.\ 2006b, \apjl, 646, L131
\bibitem[Fryer et al.(2006c)]{fryer06} Fryer, C.~L., Young, P.~A., \& Hungerford, A.~L.\ 2006c, \apj, 650, 1028
\bibitem[Fryer \& Young(2007)]{fryer=young07} Fryer, C.~L., \& Young, P.~A.\ 2007, \apj, 659, 1438
\bibitem[Fryer et al.(2007)]{fryer07} Fryer, C.~L., Hungerford, A.~L., \& Young, P.~A.\ 2007, \apjl, 662, L55
\bibitem[Fryer(2009)]{fryer09} Fryer, C.~L.\ 2009, \apj, 699, 409
\bibitem[Fryer et al.(2009)]{fryer...09} Fryer, C.~L., Brown, P.~J., Bufano, F., et al.\ 2009, \apj, 707, 193 
\bibitem[Fryer et al.(2012)]{fryer12} Fryer, C.~L., Belczynski, K., Wiktorowicz, G., et al.\ 2012, \apj, 749, 91 
\bibitem[Fryer et al.(2013)]{fryer...13} Fryer, C.~L., Belczynski, K., Berger, E., et al.\ 2013, \apj, 764, 181

\bibitem[Gal-Yam(2012)]{galyam12} Gal-Yam, A.\ 2012, Science, 337, 927 
\bibitem[Gr{\"a}fener et al.(2011)]{grafener...11} Gr{\"a}fener, G., Vink, J.~S., de Koter, A., \& Langer, N.\ 2011, \aap, 535, A56

\bibitem[Hanke et al.(2013)]{hanke...13} Hanke, F., M{\"u}ller, B., Wongwathanarat, A., Marek, A., \& Janka, H.-T.\ 2013, \apj, 770, 66 
\bibitem[Heger et al.(2000)]{heger...00} Heger, A., Langer, N., \& Woosley, S.~E.\ 2000, \apj, 528, 368
\bibitem[Heger et al.(2003)]{heger03} Heger, A., Fryer, C.~L., Woosley, S.~E., Langer, N., \& Hartmann, D.~H.\ 2003, \apj, 591, 288
\bibitem[Herant et al.(1994)]{herant...94} Herant, M., Benz, W., Hix, W.~R., Fryer, C.~L., \& Colgate, S.~A.\ 1994, \apj, 435, 339
\bibitem[Herwig et al.(1997)]{herwig...97} Herwig, F., Bloecker, T., Schoenberner, D., \& El Eid, M.\ 1997, \aap, 324, L81 
\bibitem[Hungerford et al.(2005)]{hungerford...05} Hungerford, A.~L., Fryer, C.~L., \& Rockefeller, G.\ 2005, \apj, 635, 487

\bibitem[Janka(2012)]{janka12} Janka, H.-T.\ 2012, Annual Review of Nuclear and Particle Science, 62, 407 

\bibitem[Langer(1991)]{langer91} Langer, N.\ 1991, \aap, 252, 669
\bibitem[Langer(2012)]{langer12} Langer, N.\ 2012, \araa, 50, 107
\bibitem[Lindner et al.(2012)]{lindner...12} Lindner, C.~C., Milosavljevi{\'c}, M., Shen, R., \& Kumar, P.\ 2012, \apj, 750, 163 

\bibitem[Kaper et al.(2006)]{kaper...06} Kaper, L., van der Meer, A., van Kerkwijk, M., \& van den Heuvel, E.\ 2006, The Messenger, 126, 27
\bibitem[Kiziltan et al.(2013)]{kiziltan...13} Kiziltan, B., Kottas, A., De Yoreo, M., \& Thorsett, S.~E.\ 2013, \apj, 778, 66
\bibitem[Kochanek(2013)]{kochanek13} Kochanek, C.~S.\ 2013, arXiv:1308.0013
\bibitem[Kreidberg et al.(2012)]{kreidberg...12} Kreidberg, L., Bailyn, C.~D., Farr, W.~M., \& Kalogera, V.\ 2012, \apj, 757, 36

\bibitem[MacFadyen \& Woosley(1999)]{macfadyen=woosley99} MacFadyen, A.~I., \& Woosley, S.~E.\ 1999, \apj, 524, 262 
\bibitem[MacFadyen et al.(2001)]{macfadyen01} MacFadyen, A.~I., Woosley, S.~E., \& Heger, A.\ 2001, \apj, 550, 410
\bibitem[Maeder \& Meynet(2000)]{maeder=meynet00} Maeder, A., \& Meynet, G.\ 2000, \araa, 38, 143
\bibitem[Maeder(2009)]{maeder09book} Maeder, A. (ed.)\ 2009, Physics, Formation and Evolution of Rotating Stars (Berlin: Springer)
\bibitem[Mauron \& Josselin(2011)]{mauron=josselin11} Mauron, N., \& Josselin, E.\ 2011, \aap, 526, A156
\bibitem[Milosavljevi{\'c} et al.(2012)]{milosavljevic...12} Milosavljevi{\'c}, M., Lindner, C.~C., Shen, R., \& Kumar, P.\ 2012, \apj, 744, 103
\bibitem[Moffat(2008)]{moffat08} Moffat, A.~F.~J.\ 2008, Clumping in Hot-Star Winds, ed. W.-R. Hamann, A. Feldmeier, \& L. M. Oskinova (Potsdam, Germany: Univ. Press), 17
\bibitem[Mokiem et al.(2007)]{mokiem...07} Mokiem, M.~R., de Koter, A., Vink, J.~S., et al.\ 2007, \aap, 473, 603
\bibitem[Moriya et al.(2010)]{moriya...10} Moriya, T., Tominaga, N., Tanaka, M., et al.\ 2010, \apj, 719, 1445

\bibitem[Nice et al.(2008)]{nice...08} Nice, D.~J., Stairs, I.~H., \& Kasian, L.~E.\ 2008, 40 Years of Pulsars: Millisecond Pulsars, Magnetars and More, 983, 453

\bibitem[O'Connor \& Ott(2011)]{oconner11} O'Connor, E., \& Ott, C.~D.\ 2011, \apj, 730, 70
\bibitem[Ofek et al.(2013)]{ofek...13} Ofek, E.~O., Zoglauer, A., Boggs, S.~E., et al.\ 2013, arXiv:1307.2247
\bibitem[{\"O}zel et al.(2010)]{ozel...10} {\"O}zel, F., Psaltis, D., Narayan, R., \& McClintock, J.~E.\ 2010, \apj, 725, 1918
\bibitem[{\"O}zel et al.(2012)]{ozel...12} {\"O}zel, F., Psaltis, D., Narayan, R., \& Santos Villarreal, A.\ 2012, \apj, 757, 55

\bibitem[Popham et al.(1999)]{popham99} Popham, R., Woosley, S.~E., \& Fryer, C.\ 1999, \apj, 518, 356

\bibitem[Quimby et al.(2011)]{quimby...11} Quimby, R.~M., Kulkarni, S.~R., Kasliwal, M.~M., et al.\ 2011, \nat, 474, 487 

\bibitem[Scheck et al.(2004)]{scheck...04} Scheck, L., Plewa, T., Janka, H.-T., Kifonidis, K., \& M\"uller, E.\ 2004, Physical Review Letters, 92, 011103
\bibitem[Schwab et al.(2010)]{schwab...10} Schwab, J., Podsiadlowski, P., \& Rappaport, S.\ 2010, \apj, 719, 722
\bibitem[Sedov(1959)]{sedov59} Sedov, L.~I.\ 1959, Similarity and Dimensional Methods in Mechanics, New York: Academic Press, 1959
\bibitem[Stothers \& Chin(1985)]{stothers=chin85} Stothers, R.~B., \& Chin, C.-W.\ 1985, \apj, 292, 222
\bibitem[Surman et al.(2011)]{surman11} Surman, R., McLaughlin, G.~C., \& Sabbatino, N.\ 2011, \apj, 743, 155

\bibitem[Thorsett \& Chakrabarty(1999)]{thorsett=chakrabarty99} Thorsett, S.~E., \& Chakrabarty, D.\ 1999, \apj, 512, 288
\bibitem[Timmes et al.(1996)]{timmes96} Timmes, F.~X., Woosley, S.~E., Hartmann, D.~H., \& Hoffman, R.~D.\ 1996, \apj, 464, 332

\bibitem[Ugliano et al.(2012)]{ugliano12} Ugliano, M., Janka, H.-T., Marek, A., \& Arcones, A.\ 2012, \apj, 757, 69 

\bibitem[Vink et al.(2011)]{vink...11} Vink, J.~S., Muijres, L.~E., Anthonisse, B., et al.\ 2011, \aap, 531, A132

\bibitem[Warren \& Salmon(1993)]{warren&salmon93} Warren, M.~S., \& Salmon, J.~K.\ 1993,
in Supercomputing '93, ed. IEEE Computer Society (Los Alamitos; IEEE Comput. Soc.), 12
\bibitem[Warren \& Salmon(1995)]{warren&salmon95} Warren, M.~S., \& Salmon, J.~K.\ 1995, Computer Physics Communications, 87, 266
\bibitem[Weaver et al.(1978)]{weaver...78} Weaver, T.~A.,  Zimmerman, G.~B., \& Woosley, S.~E.\ 1978, \apj, 225, 1021
\bibitem[Woosley(1989)]{woosley89} Woosley, S.~E.\ 1989, Annals of the New York Academy of Sciences, 571, 397
\bibitem[Woosley(1993)]{woosley93} Woosley, S.~E.\ 1993, \apj, 405, 273 
\bibitem[Woosley \& Weaver(1995)]{woosley=weaver95} Woosley, S.~E., \& Weaver, T.~A.\ 1995, \apjs, 101, 181 
\bibitem[Woosley et al.(2002)]{woosley...02} Woosley, S.~E., Heger, A., \& Weaver, T.~A.\ 2002, Reviews of Modern Physics, 74, 1015

\bibitem[Yoon et al.(2006)]{yoon...06} Yoon, S.-C., Langer, N., \& Norman, C.\ 2006, \aap, 460, 199
\bibitem[Young \& Arnett(2005)]{young=arnett05} Young, P.~A., \& Arnett, D.\ 2005, \apj, 618, 908
\bibitem[Young \& Fryer(2007)]{young07} Young, P.~A., \& Fryer, C.~L.\ 2007, \apj, 664, 1033 

\bibitem[Zhang \& Fryer(2001)]{zhang&fryer01} Zhang, W., \& Fryer, C.~L.\ 2001, \apj, 550, 357
\bibitem[Zhang et al.(2008)]{zhang08} Zhang, W., Woosley, S.~E., \& Heger, A.\ 2008, \apj, 679, 639 

\end{thebibliography}
\end{document}